\documentclass[11pt,a4paper]{article}
\pdfoutput=1
\usepackage{jheppub}
\usepackage{amsmath}
\usepackage{verbatim}
\usepackage{graphicx}
\usepackage{mathrsfs}
\usepackage{appendix}
\usepackage{caption}
\usepackage{float}
\usepackage{subfig}
\usepackage{mathtools}
\usepackage{mathtools}

\usepackage{amssymb}
\usepackage{inputenc, array}
\usepackage{textcomp}
\usepackage[dvipsnames]{xcolor}

\newcommand{\e}{\epsilon}

\renewcommand{\L}{{\mathcal{L}}}
\newcommand{\bcL}{\bar{{\mathcal{L}}}}

\newcommand{\h}{{\bar h}}

\newcommand{\be}[1]{ \begin{equation}\label{#1} }
\newcommand{\ee}{\end{equation}}
\newcommand{\bea}[1]{\begin{eqnarray}\label{#1} }
\newcommand{\eea}{\end{eqnarray}}
\newcommand{\bes}{\begin{subequations}}
\newcommand{\ees}{\end{subequations}}

\newcommand{\p}{\partial}
\renewcommand{\a}{\alpha}
\renewcommand{\b}{\beta}
\renewcommand{\t}{\tau}

\newcommand{\s}{\sigma}

\newcommand{\refb}[1]{(\ref{#1})}

\renewcommand{\>}{\rangle}

\DeclarePairedDelimiterX\braket[2]{\langle}{\rangle}{#1 \delimsize\vert #2}

\newcommand{\ie}{\emph{i.e.}}

\renewcommand{\a}{\alpha}

\renewcommand{\b}{\beta}
\renewcommand{\t}{\tau}

\title{BMS Field Theories and Weyl Anomaly}

\author{Arjun Bagchi,} \author{Sudipta Dutta,} \author{Kedar S.~Kolekar,} \author{and Punit Sharma. \\}

\affiliation{Indian Institute of Technology Kanpur, Kalyanpur, Kanpur 208016. INDIA. \\}

\emailAdd{(abagchi, dsudipta, kedarsk, spunit)@iitk.ac.in}

\abstract{Two dimensional field theories with Bondi-Metzner-Sachs symmetry have been proposed as duals to asymptotically flat spacetimes in three dimensions. These field theories are naturally defined on null surfaces and hence are conformal cousins of Carrollian theories, where the speed of light goes to zero. In this paper, we initiate an investigation of anomalies in these field theories. Specifically, we focus on the BMS equivalent of Weyl invariance and its breakdown in these field theories and derive an expression for Weyl anomaly. Considering the transformation of partition functions under this symmetry, we derive a Carrollian Liouville action different from ones obtained in the literature earlier. }

\preprint{}

\begin{document}
\maketitle

\vfill

\section{Introduction}
The use of symmetry is the most basic and most useful of tools available to a theoretical physicist. All symmetries of a classical physical system do not automatically graduate to quantum mechanical symmetries of the corresponding quantum system. These are called anomalies. Consider a physical system governed by a classical action $S_{cl}$. $S_{cl}$ is invariant under a symmetry group $G$, i.e. $\delta_G S_{cl} =0$. If the quantum mechanical action $S_{q}$, obtained by quantising the classical system, does not respect symmetry under $G$ ($\delta_G S_{q} \neq0$), then $G$ is said to be anomalous. Anomalies can be discrete or continuous, and anomalies can arise in global or gauge symmetries. 
Global anomalies can have interesting physical consequences. These may mean e.g. that classical selection rules are not obeyed in the quantized theory and that some processes disallowed by the classical analysis may actually take place in the quantum world. Anomalous global symmetries are also very useful tools for non-pertubative analysis of quantum field theories. Anomalous gauge symmetries, on the other hand, are indicators of sickness and need to be canceled in order to have a physically consistent theory. Anomalies have had a rich history and we point to the excellent reviews \cite{Harvey:2005it,Bilal:2008qx} for further details. 

The discussion of anomalies in the past, with due justification, have principally focussed on relativistic quantum field theories. In more recent times, there has been work on anomalies in the context of non-relativistic systems \cite{Jensen:2014hqa,Arav:2014goa,Auzzi:2015fgg,Arav:2016xjc,Pal:2016rpz,Auzzi:2016lrq,Auzzi:2017jry}. In our paper, we initiate a discussion of anomalies for the case of field theories living on null hypersurfaces, also called Carrollian field theories{\footnote{See, however, \cite{Jensen:2017tnb} for a discussion of anomalies in 2d Warped CFTs. These turn out to be Carrollian CFTs for scaling exponent $z=0$. We are however interested in $z=1$ theories, that can be obtained as an ultra-relativistic limit of 2d CFTs, as we go on to explain in the paper.}}. Our explorations in this paper will be limited to conformal cousins of Carrollian field theories and the question of breaking of conformal invariance at a quantum level. In other words, we will discuss Weyl symmetry and Weyl anomaly in the context of two-dimensional field theories living on null surfaces. 

\subsection*{Weyl, Carroll, BMS}

Classically, Weyl invariance of a system is the invariance of the classical action under a local rescaling of the metric
\be{weyl}
g_{\mu\nu}(x) \mapsto e^{2 \Omega(x)} g_{\mu \nu}(x).
\ee
For generally covariant theories, conformal invariance is a direct consequence of Weyl invariance. Conformal transformations are the ones that change the metric upto some Weyl scaling. Hence for theories  that are Weyl invariant, conformal symmetry would arise as residual symmetries after fixing a background choice. A direct consequence of Weyl invariance is the vanishing of the trace of the energy-momentum tensor of a field theory. Weyl symmetry is anomalous in the quantum regime and this leads to a breaking of conformal symmetry. In two dimensions, where the underlying symmetry of conformal field theories enhance to two copies the Virasoro algebra, the Weyl anomaly is proportional to the central charge multiplied by the Ricci scalar of the manifold the CFT lives on. We will review all of this basically textbook material briefly in the next section. 

\medskip 

\noindent Our focus in this paper is on theories which are Carrollian instead of Poincare invariant. The wonderfully named Carroll group (after Lewis Carroll of {\em Alice's Adventures in Wonderland} fame) is obtained by a priori a rather bizarre contraction of the Poincare group, where the speed of light goes to zero \cite{LevyLeblond}. Interestingly, these theories encompass theories living on null hypersurfaces. Two of the most important classes of null hypersurfaces that we encounter in physics are the null boundary of asymptotically flat spacetimes and the event horizons of generic black holes. If we are interested in defining field theories that live on null surfaces, it is of interest that the degrees of freedom of the field theory don't leave this null surface at a later time. Any massive degree of freedom cannot travel at the speed of light and hence would not do the job. So it is natural to consider massless Carrollian theories, which are tantamount to conformal Carrollian theories.

\medskip 

\noindent For asymptotically flat spacetimes, the asymptotic symmetry groups at null infinity are the Bondi-Metzner-Sachs (BMS) groups \cite{Bondi:1962px,Sachs:1962zza}, which in dimensions three \cite{Barnich:2006av} and four are infinite dimensional groups. After a long hiatus since its original discovery in the context of four dimensional spacetimes in the late 1960s, the theoretical physics community has now woken up to the realisation that the BMS group should be omnipresent in the discussions of scattering theory in quantum field theories defined in asymptotically flat spacetimes. Following the lead of Strominger \cite{Strominger:2013jfa}, a large volume of literature has built up linking the BMS group with soft theorems and memory effects in a triangle of relations. We refer the interested reader to \cite{Strominger:2017zoo} for a detailed discussion on these issues and related references. 

\medskip 

\noindent The conformal Carrollian field theories that we are discussing in this work are also BMS-invariant field theories. This stems from the isomorphism between these groups \cite{Duval:2014lpa,Duval:2014uoa,Duval:2014uva}. This isomorphism was discovered following an even more startling observation of the isomorphism between Galilean conformal algebras and the BMS algebra \cite{Bagchi:2010zz}. Field theories defined on these Carrollian backgrounds suffer from the absence of a non-degenerate metric; the theories are defined on a fibre-bundle structure rather than a (pseudo-) Riemmannian one. Our task of investigating Weyl symmetry would thus be complicated by this feature. In what follows, we shall be exclusively focussing on conformal Carrollian or BMS invariant theories in 2 spacetime dimension. 

\subsection*{BMS Applications: Flat Holography and Tensionless Strings} 
For the sake of the uninitiated reader, let us pause to give a bit of context to why we are interested in investigating theories that seem very exotic at the outset. 
\medskip 

\noindent{\em{Holography in Flat Spacetimes}}

\medskip

\noindent 
The Holographic Principle \cite{tHooft:1993dmi,Susskind:1994vu} has changed the way we look at quantum gravity and AdS/CFT \cite{Maldacena:1997re} is by far the most impressive of our present tools to attack this age-old problem of bringing quantum mechanics and gravity together. Although there has been spectacular progress in understanding Anti de Sitter spacetime and its dual field theories, particularly in the context of the original Maldacena correspondence between string theory on AdS$_5 \times$ S$^5$ and $\mathcal{N}=4$ Super Yang Mills theory in $d=4$, progress has surprisingly been much less impressive in non-AdS scenarios. The question of holography for asymptotically flat spacetimes, a subject that is clearly more relevant to the real world, especially for applications in say astrophysics and astrophysical black holes, has been receiving some attention more recently. Some of the earlier works include \cite{Bagchi:2010zz,deBoer:2003vf,Barnich:2009se,Barnich:2010eb}. More recent attempts based on holography of the celestial sphere have built up following Strominger et al. These are somewhat tangential to our work here and we shall refer the reader to \cite{Strominger:2017zoo} again for a more detailed exposition.  
\medskip 

\noindent 
In 3d asymptotically flat bulk and two dimensional boundary theory, progress has been somewhat better. The symmetry group in question is BMS$_3$. Following the assertion that the putative dual theory has to be a BMS-invariant 2d field theory in \cite{Bagchi:2010zz,Bagchi:2012cy}, a body of work has emerged, a selected, non-exhaustive list of which includes \cite{Bagchi:2012xr}--\cite{Ciambelli:2018wre}. The study of 2d BMS invariant field theories is thus important to building on this proposed correspondence. 
\medskip 

\noindent{\em{Dual theories to generic black holes}}

\medskip

\noindent 
As we remarked earlier, Carrollian structures appear on generic null surfaces. Hence if we are looking at building field theories which live on  generic black hole horizons, it is natural to consider Carrollian theories. (See \cite{Donnay:2019jiz} for a detailed analysis of the emergence of Carrollian structures on black hole event horizons.) Also, as per our previous argument, if we want degrees of freedom of the theories not to leave the null surface, we must necessarily consider massless or conformal Carrollian theories. There have been attempts at understanding these horizon BMS symmetries in recent times e.g. in \cite{Hawking:2016msc}. BMS-invariant field theories thus can act as putative duals for generic black holes. A derivation of the entropy of black holes based on the BMS-Cardy formula \cite{Bagchi:2012xr} can be found in \cite{Carlip:2017xne,Carlip:2019dbu}.

\medskip 

\noindent{\em{The theory of tensionless strings}}

\medskip

\noindent Understanding the ultra-high energy regime of string theory has been long sought-after, especially after the work of Gross and Mende \cite{Gross:1987ar,Gross:1987kza} who found substantial simplifications in string scattering in this domain. This is the regime explored by the tensionless limit of string theory, where the worldsheet of the string becomes null and the worldsheet metric becomes degenerate. In this limit, the worldsheet conformal symmetry is replaced by BMS$_3$, which now arises as residual symmetries of the tensionless worldsheet \cite{Isberg:1993av,Bagchi:2013bga}. So the theory of tensionless strings is organised by this underlying BMS algebra in the same way as usual string theory is dictated by worldsheet conformal invariance \cite{Bagchi:2015nca}. This is thus another very important place where two dimensional BMS invariant theories play a pivotal role {\footnote{It is of interest to note here that some rather unique phenomena in the quantum tensionless strings have been recently unearthed \cite{Bagchi:2020ats, Bagchi:2020fpr, Bagchi:2019cay}, a lot of which is based on a better understanding of the underlying BMS algebra.}}. It is important to stress here that the BMS symmetries that arise here are gauge symmetries, as opposed to the case of putative dual theories to flatspace, where BMS symmetries are global symmetries of the field theory.  

\smallskip

\noindent{\em{Galilean CFTs}}

\smallskip

\noindent As mentioned above, there is a rather startling duality between non-relativistic systems (where speed of light $c\to \infty$) and ultra-relativistic or Carrollian systems ($c\to 0$) in spacetime dimensions $d=2$ \cite{Bagchi:2010zz}. In general dimensions, both systems have degenerate metrics and hence non-Riemannian background structures. These structures are fibre bundles. In the case of non-relativistic systems, the base of the fibre bundle is the time direction ${\rm I\!R}_t$ and the fibres are $(d-1)$ dimensional spatial slices ${\rm I\!R}^{d-1}$ \cite{Bagchi:2009my, Duval:2009vt}. The Carrollian limit interchanges the base and the fibre \cite{Duval:2014uoa}. In the case of $d=2$, the base and fibre are both one dimensional and interchanging them does not change the background symmetry algebra. So Carrollian and Galilean algebras in $d=2$ are isomorphic and the isomorphism extends to their conformal counterparts, the Carrollian conformal and the Galilean conformal algebras. Hence, results constructed for 2d Carrollian CFTs are valid (upto an exchange of spatial and temporal directions for coordinate dependent answers) for 2d Galilean CFTs and vice-versa.

\subsection*{Weyl anomaly and BMS} 
The Weyl anomaly has had a very rich history dating back almost five decades{\footnote{For a wonderful memoir of the development of Weyl anomaly, the reader is pointed to  \cite{Duff:1993wm}.}} and has found applications in diverse fields such as quantum gravity, black hole physics, inflationary cosmology, string theory and statistical physics. We are now concerned about the form of Weyl symmetry and Weyl anomaly in 2d BMS-invariant field theories, which as stated above, would be useful for understanding aspects of holography in asymptotically flat 3d spacetime, or for tensionless string theories. It is possible that this construction would find similar applications in logarithmic corrections to entropies of Flat space cosmologies in 3d \cite{Bagchi:2013qva}, aspects of inflationary cosmology, and even statistical physics. Our constructions of the Weyl symmetry and its anomaly in this paper would be purely from the field theoretic side. One can envision that a holographic computation in flat spacetimes, following the lead of \cite{Henningson:1998gx} and with suitable modifications for the asymptotically flat case, would reproduce the answers we obtain in this paper. 

\subsection*{Outline for the rest of the paper}
A brief outline of the paper is as follows. In Sec.~2, we revisit the well-known features of Weyl invariance and Weyl anomaly in the context of 2d CFTs, in an effort to set notation and set the stage for our explorations of Weyl symmetry in BMS invariant field theories.  In Sec.~3, we give a brief review of the BMS algebra and 2d field theories invariant under the BMS$_3$ algebra, which we will call BMS field theories or BMSFTs for short. We review some basic notions that we would require for calculations with 2d BMSFTs later in the paper. In Sec.~4,  we introduce TT OPEs and delta functions on our Carrollian backgrounds and show how the delta function leads to Ward identities and correlation functions which match and generalise answers in \cite{Bagchi:2015wna}. Armed with these formulae, we compute the BMS Weyl anomaly in Sec.~5. We conclude with a summary and discussions about current and future directions of further research in Sec.~6. A number of appendices contain more details about our computations that are omitted in the main body of the paper.

\section{A quick recap: Conformal symmetry and Weyl anomaly}\label{sec:2DCFT}

As is very well known, the power of conformal symmetry in two dimensions is greatly enhanced by the underlying infinite dimensional Virasoro algebra (two copies of it):
\begin{eqnarray}
&& [\mathcal{L}_n,\mathcal{L}_m] =(n-m)\mathcal{L}_{n+m}+\frac{c}{12}( m^3-m)\delta_{n+m, 0}, \nonumber \\
&& [\mathcal{\bar{L}}_n,\mathcal{\bar{L}}_m] =(n-m)\mathcal{\bar{L}}_{n+m}+\frac{\bar{c}}{12}(m^3-m)\delta_{n+m, 0},  \label{Virasoro}  \\
&& [\mathcal{L}_n,\mathcal{\bar{L}}_m] =0. \nonumber
\end{eqnarray}
Here $\mathcal{L}_n$ and $\mathcal{\bar{L}}_n$ are the holomorphic and anti-holomorphic generators of the algebra respectively, whereas $c$ and $\bar{c}$ are the central charges. For the purposes of this section, we shall take $c=\bar{c}$. We present a brief discussion of the $c\neq\bar{c}$ case at the end of the section. 

\medskip

\noindent We now investigate Weyl invariance and its implications on the stress energy tensor. Consider a field theory coupled to some non-trivial background described by the action
\begin{equation}
	S[\Phi, g_{\mu\nu}]=\int d^2 x \sqrt{-g} \, \mathcal{L}[\Phi,g_{\mu\nu}].
\end{equation}
Variation of the action with respect to the metric is 
\be{}
\delta S[\Phi, g_{\mu\nu}]= \frac{1}{2\pi} \int d^2 x \sqrt{-g} \, \, T_{\mu\nu}\delta g^{\mu \nu},\quad \text{where} \quad T_{\mu\nu} = \frac{2\pi}{\sqrt{-g}}\frac{\delta S}{\delta g^{\mu\nu}}
\ee
is the energy-momentum (EM) tensor. Under Weyl transformation \refb{weyl}, the metric changes by an overall scale factor. For an infinitesimal Weyl transformation, this is just $\delta g_{\mu \nu} = 2\Omega(x) g_{\mu\nu}.$ The variation of the action under the infinitesimal Weyl scaling  becomes 
\begin{equation}
\delta S[\Phi, g_{\mu\nu}] = -\frac{1}{2\pi} \int d^2 x \sqrt{-g}\,2 \,\Omega(x)T^{\mu}_{\ \mu}.
\end{equation}
Thus, we see that field theories invariant under Weyl transformations have $T^{\mu}_{\ \mu}=0$. This vanishing of trace of EM tensor  is a defining feature of conformal field theories.

\medskip

\noindent Although the above statement defines CFTs at classical level, the trace of EM tensor suffers from an anomaly and  no longer vanishes in quantum theories. This is because of the presence of central charge in Virasoro algebra. The central charge captures the effect of soft breaking of conformal symmetry in the presence  of a scale. When the theory lives on a curved background, the curvature introduces a scale  in the theory. In $d=2$, the presence of the central charge $c$ in the algebra leads to the trace anomaly
\begin{equation}\label{cft-trace}
\langle T^{\mu}_{\ \mu} \rangle = \frac{c}{12} \mathcal{R}.
\end{equation}
Here $\mathcal{R}$ is the Ricci scalar of the background manifold. Below we derive this result. 

\subsection*{Derivation of trace anomaly}
To set the stage for our calculations for the Weyl or trace anomalies in BMS invariant field theories later in the paper, we now briefly review a derivation of the same in 2d CFTs \cite{Tong:2009np}. We will do this for CFTs on backgrounds that are infinitesimally close to flat space. For this purpose, consider a CFT on a complex plane parametrized by coordinates $(z,\bar{z})$. The Operator Product Expansion (OPE) of the holomorphic $(T(z)\equiv T_{zz}(z))$ and antiholomorphic $(\bar{T}(\bar{z})\equiv T_{\bar{z}\bar{z}}(\bar{z}))$ parts of EM tensor of a 2d CFT are given by
\bea{cft:TT-ope}
T(z) T(z') \sim \frac{c}{2(z-z')^4} + \frac{2 T(z')}{(z-z')^2} + \frac{\partial' T(z')}{(z-z')} + \text{reg} \cr
\bar{T}(\bar{z}) \bar{T}(\bar{z}') \sim \frac{\bar{c}}{2(\bar{z}-\bar{z}')^4} + \frac{2 \bar{T}(\bar{z}')}{(\bar{z}-\bar{z}')^2} + \frac{\bar{\partial}' \bar{T}(\bar{z}')}{(\bar{z}-\bar{z}')} + \text{reg}
\eea
where `reg' denotes regular terms as $(z,\bar{z}) \rightarrow (z',\bar{z}')$. We will adopt the following shorthands for the off-diagonal components of the EM tensor
\begin{equation}
\mathcal{T}\equiv T_{\bar{z}z}(z,\bar{z}), \quad  \bar{\mathcal{T}}\equiv T_{z\bar{z}}(z,\bar{z}),
\end{equation}
The components of the conservation equation $\partial^{\mu}T_{\mu\nu} = 0$ are
\begin{equation}
	\bar{\partial}T + \partial \mathcal{T} = 0, \qquad \partial \bar{T} + \bar{\partial}\bar{\mathcal{T}} = 0,
\end{equation}
Using the first of the above equations and $TT$ OPE, we can write
\begin{equation}
	\partial\partial' \mathcal{T}\mathcal{T}' = \bar{\partial}\bar{\partial}' T T' = \bar{\partial}\bar{\partial}' \Big(\frac{c}{2(z-z')^4} + \cdots \Big) = \bar{\partial}\bar{\partial}' \Big(\frac{c}{12}\partial^2\partial'\Big(\frac{1}{z-z'}\Big)\Big).
\end{equation}
Here $\mathcal{T}'$ is a shorthand for writing $\mathcal{T}(z',\bar{z}')$.
Using the delta function 
\be{delta-r}\bar{\partial}\frac{1}{z-z'} = 2\pi \delta^{(2)}(z-z',\bar{z}-\bar{z}'),
\ee 
the above equation becomes
\begin{equation}
	\partial\partial' \mathcal{T}\mathcal{T}' = \frac{\pi c}{6} \bar{\partial}'\partial^2\partial'\delta^{(2)}(z-z',\bar{z}-\bar{z}'),
\end{equation}
which implies
\begin{equation}
	\mathcal{T}\mathcal{T}' = \frac{\pi c}{6} \bar{\partial}'\partial\delta^{(2)}(z-z',\bar{z}-\bar{z}')\ .
\end{equation}
Similarly using the second conservation equation, $\bar{T}\bar{T}$ OPE and an $\delta$-function identity similar to \refb{delta-r}, we get
\begin{equation}
	\bar{\mathcal{T}}\bar{\mathcal{T}}' = \frac{\pi c}{6} \partial'\bar{\partial}\delta^{(2)}(z-z',\bar{z}-\bar{z}').
\end{equation}
Using the conservation equations and $T\bar{T}$ OPE we can also show that $\mathcal{T}\bar{\mathcal{T}}' \sim \text{reg}$. Then the OPE of trace ($T^{\mu}_{\ \mu} = 2(\mathcal{T} + \bar{\mathcal{T}})$) is
\begin{eqnarray}
	T^{\mu}_{\ \mu}T'^{\nu}_{\ \nu} = 4(\mathcal{T}\mathcal{T}' + \bar{\mathcal{T}}\bar{\mathcal{T}}') = \frac{2\pi c}{3}\Big(\bar{\partial}'\partial\delta^{(2)}(z-z',\bar{z}-\bar{z}') + \partial'\bar{\partial}\delta^{(2)}(z-z',\bar{z}-\bar{z}') \Big). \label{cft-trace-ope}
\end{eqnarray}
Now consider an infinitesimal Weyl transformation of the plane metric, \ie\ $\delta g_{\mu\nu} = 2\Omega \delta_{\mu\nu}$, where $\Omega(z,\bar{z})$ is small. We can expand $\langle T^{\mu}_{\ \mu}\rangle_{g}$ in $\Omega$ as
\begin{equation}
	\langle T^{\mu}_{\ \mu}\rangle_{g} = \langle T^{\mu}_{\ \mu}\rangle_{\delta} +\delta \langle T^{\mu}_{\ \mu}\rangle_{\delta} + (\text{higher order in}\ \Omega), \label{cft:trace-expansion}
\end{equation}
where the subscript `$g$' denotes that the expectation value is evaluated in a CFT on a background metric $g_{\mu\nu}$. From the definition, $\langle T^{\mu}_{\ \mu}\rangle_{\delta} = \frac{1}{Z}\int \mathcal{D}\Phi e^{-S}T^{\mu}_{\ \mu}$, we can compute the change in $\langle T^{\mu}_{\ \mu}\rangle_{\delta}$ under an infinitesimal Weyl transformation $\delta g_{\mu\nu} = 2\Omega \delta_{\mu\nu}$ as
\begin{equation}
	\delta \langle T^{\mu}_{\ \mu}\rangle_{\delta} = \frac{1}{Z}\int \mathcal{D}\Phi e^{-S} \Big(-\frac{\delta S}{\delta g^{\alpha\beta}}\delta g^{\alpha\beta} T^{\mu}_{\ \mu} \Big) = \frac{1}{Z}\int \mathcal{D}\Phi e^{-S} \int \frac{d^2z'}{2\pi} T^{\mu}_{\ \mu}T'^{\nu}_{\ \nu}\Omega'.
\end{equation}
Substituting the $T^{\mu}_{\ \mu}T'^{\nu}_{\ \nu}$ OPE \eqref{cft-trace-ope} and simplifying, we get
\begin{eqnarray}
	\delta \langle T^{\mu}_{\ \mu}\rangle_{\delta} =  \frac{2\pi c}{3} \int d^2z'\Big(\bar{\partial}'\partial\delta^{(2)}(z-z',\bar{z}-\bar{z}') + \partial'\bar{\partial}\delta^{(2)}(z-z',\bar{z}-\bar{z}') \Big)\Omega' = -\frac{2 c}{3} \partial\bar{\partial}\Omega.\cr \label{cft:delta T}
\end{eqnarray}
Now, since the anomaly arises from regulating short distance divergences, $\langle T^{\mu}_{\ \mu}\rangle$ is same for all states in the theory. So $\langle T^{\mu}_{\ \mu}\rangle$ is equal to some geometric quantity of the background metric, which must be local and dimension $2$, \ie\ the Ricci scalar $\mathcal{R}$. The Weyl anomaly vanishes on the plane due to translational invariance \ie\ $\langle T^{\mu}_{\ \mu}\rangle_{\delta} = 0$. This is consistent with the above argument. Now when we consider deformations away from the plane, starting with a metric of the form:  
$$g_{\mu\nu} = e^{2\Omega}\delta_{\mu\nu}  \Rightarrow \, \mathcal{R} = -8e^{-2\Omega}\partial\bar{\partial}\Omega \, \Rightarrow \, \mathcal{R} \approx -8\partial\bar{\partial}\Omega \quad \text{for infinitesimal $\Omega$}. $$ Putting all these together and \eqref{cft:delta T} in \eqref{cft:trace-expansion}, we can write the trace anomaly
\be{cft-anom}
	\langle T^{\mu}_{\ \mu}\rangle_{g} = -\frac{2 c}{3} \partial\bar{\partial}\Omega + (\text{higher order in}\ \Omega) = \frac{c}{12}\mathcal{R}.
\ee
The above is of course very well known and basically text book material. But it is important for us to keep this logical sequence in mind as we build towards a similar derivation of the Weyl anomaly, now in the context of 2d field theories governed by a different symmetry, {\it{i.e.}} the BMS$_3$ algebra. 

\subsection*{Unequal central terms}
We had assumed at the beginning of the section that we would take $c=\bar{c}$ in our derivations of the Weyl anomaly. Now let us briefly comment on the choice and what happens if we choose unequal left and right central terms. From the point of view of the 2d field theory, $c\neq\bar{c}$ implies a violation of parity in the theory. From the context of AdS$_3$/CFT$_2$, this shows up when one adds a gravitational Chern-Simons term to the Einstein-Hilbert action. The resulting theory, called Topologically Massive Gravity, is one where a diffeomorphism anomaly arises. The stress-energy tensor is not conserved any more: 
\be{}
\nabla_\mu T^{\mu\nu} = k (c-\bar{c}) g^{\mu\nu} \varepsilon^{\a\b} \p_\a \p_\s \Gamma^\s_{\mu \b}.
\ee
In the above $k$ is a constant and $k (c-\bar{c})$ is the coefficient of the diffeomorphism anomaly. This anomaly can be equivalently traded for a Lorentz anomaly:
\be{}
\Theta^{\mu\nu} - \Theta^{\nu\mu} = k (c-\bar{c}) \varepsilon^{\mu\nu}\mathcal{R}, 
\ee
where the improved stress tensor $\Theta^{\mu\nu}$ is no longer symmetric, but now does obey the conservation equation $\nabla_\mu \Theta^{\mu\nu}=0$. It is clear that our earlier simple-minded derivation of the Weyl anomaly would run into rough waters here. The expression for the Weyl anomaly actually depends on the formulation. For a stress tensor that yields consistency with Wess-Zumino conditions, the Weyl anomaly gets a correction proportional to $c-\bar{c}$. For a covariant stress tensor, constructed at the expense of not satisfying the Wess-Zumino conditions, one get the same expression as \refb{cft-anom}, with $c$ replaced by $\frac{1}{2}(c+\bar{c})$. Finally, for a non-symmetric stress tensor where we have traded the gravitational anomaly for a Lorentz anomaly, the Weyl anomaly is again the same as the covariant case, albeit that the conservation equation now holds. We point the reader to \cite{Kraus:2005zm, Skenderis:2009nt} for further details on this. 


\section{BMS field theories: a reminder of the basics}\label{sec:BMStheories}
The asymptotic symmetries of flat spacetimes in three dimensions at the null boundary $\mathscr{I}^\pm$, which is topologically ${\rm I\!R}_{\text{null}} \times S^1$, is encapsulated in the BMS$_3$ algebra:
\begin{eqnarray}\label{BMSalgebra-ce}
&& [L_n,L_m]=(n-m)L_{n+m}+\frac{c_L}{12}n(n^2-1)\delta_{n+m,0} , \nonumber \\
&& [L_n,M_m]=(n-m)M_{n+m}+\frac{c_M}{12}n(n^2-1)\delta_{n+m,0}, \\ \nonumber
&& [M_n,M_m]=0 .
\end{eqnarray}
Here $L_n$ and $M_n$ generate diffeomorphisms of the $S^1$ at $\mathscr{I}^\pm$ (superrotations) and angle dependent translations of the null direction (supertranslation) respectively. $c_L$ and $c_M$ are central terms which for Einstein gravity become $c_L=0, c_M= \frac{3}{G}$ \cite{Barnich:2006av}.

Our prescription of Minkowskian holography draws heavily on the evolution of AdS/CFT from the seminal work of Brown and Henneaux, who famously computed the asymptotic symmetries of AdS$_3$ to obtain two copies of the Virasoro algebra that have since been understood as the global symmetries of the dual 2d field theory. Following this lead, it has been proposed \cite{Bagchi:2010zz, Bagchi:2012cy} that the dual of 3d Minkowski spacetimes is a 2d field theory living on $\mathscr{I}^\pm$ which has BMS$_3$ as its underlying symmetry. We shall call these field theories BMS field theories or BMSFTs in short.  

\subsection{Null cylinder and null plane}
Since we would like a holographic understanding of Minkowski spacetimes, and we have just mentioned that the BMSFT sits on $\mathscr{I}^\pm$ which has the topology of cylinder, a convenient representation of the generators of the BMS$_3$ algebra in terms of differential operators is the so-called cylinder representation:
\be{bms-cyl}
L_n=ie^{in\sigma}(\partial_{\sigma}+in\tau\partial_{\tau}), \qquad M_n=ie^{in\sigma}\partial_{\tau}.
\ee
where $\s$ is the angular coordinate of the $S^1$ and $\t$ parametrises the null direction ${\rm I\!R}_{\text{null}}$. We will also find it particularly useful to speak of the plane representation where the field theory is defined on ${\rm I\!R}_{\text{null}} \times {\rm I\!R}$. Here the $S^1$ in the cylinder representation is unwrapped. The generators here take the form
\be{bms-gen-pl}
L_n= -x^{n+1}\partial_x-(n+1)x^{n}t\partial_t, \qquad M_n=x^{n+1}\partial_t.
\ee
The map from the null cylinder to the null plane is
\be{}
t=-i\tau e^{i\sigma}, \qquad  x=e^{i\sigma}. 
\ee
The energy momentum tensors of the 2d BMSFT can be defined through the generators \cite{Bagchi:2010vw}:
\be{}
T_1(x, t) = \sum_{n}\Big[L_n+(n+2)\frac{t}{x}M_n\Big]x^{-n-2}, \qquad T_2 (x, t)= \sum_{n} M_n x^{-n-2}.
\ee
One can also find the expressions of the EM tensor on the cylinder and these are related to each other by the BMS version of the Schwartzian derivative \cite{Basu:2015evh}: 
\be{}
T_1(\s, \t) =\sum_{n} (L_n-in\tau M_n)e^{-in\sigma} + \frac{c_L}{24}, \quad T_2(\s, \t) = \sum_{n} M_ne^{-in\sigma} + \frac{c_M}{24}.
\ee
We would be interested in building representations of the BMS algebra in analogy with the Virasoro algebra and hence would construct highest weight representations. The states of the theory are labeled by quantum numbers that are the eigenvalues of $L_0$ and $M_0$ :
\begin{equation}
	L_0 |h_L,h_M\rangle = h_L |h_L,h_M \rangle, \qquad  M_0|h_L,h_M\rangle= h_M |h_L,h_M\rangle.
\end{equation}
As is evident from the algebra, the operators $L_n$ and $M_n$ for $n>0$ lower the $h_L$ eigenvalue of a state. We will demand that this spectrum is bounded from below. This leads to a definition of a primary state $(|h_L,h_M\rangle_p)$: 
\begin{equation}
	L_n |h_L,h_M\rangle_p= 0 , \qquad  M_n|h_L,h_M\rangle_p= 0, \qquad \forall\ n > 0.
\end{equation}
Towers of states called the BMS modules can be built by the action of the raising operators $L_{-n}$ and $M_{-n}$ on each of the primary states of the spectrum. 

\subsection{BMS from CFT}
Asymptotically flat spacetimes can be obtained from asymptotically AdS spacetimes by taking the radius of AdS to infinity. It is thus expected that the symmetries map into each other in this singular limit. At the level of the asymptotic symmetry algebras, this limit is perceived as an In{\"o}n{\"u}-Wigner contraction: 
\be{cft2bms-ur}
L_{n}=\mathcal{L}_n-\mathcal{\bar{L}}_{-n}, \qquad  M_n=\epsilon(\mathcal{L}_n+\mathcal{\bar{L}}_{-n}),
\ee
where $\L$ and $\bcL$ represent the left and right moving Virasoros of the 2d CFT. One can think of this limit in terms of the spacetime coordinates. The boundary topology of global AdS is a cylinder and the limit takes us to the null boundary of flat space $\mathscr{I}^\pm$ which as we discussed earlier is a null cylinder ${\rm I\!R}_{\text{null}} \times S^1$. The limit can thus be thought of as an ultra-relativistic (UR) boost. In terms of spacetime coordinates, if we start with the Virasoro generators on the cylinder 
\be{vir-cyl}
\L_n = e^{inw}\partial_w, \quad \bcL_n =e^{in\bar{w}}\partial_{\bar{w}},
\ee
where $w = \tau + \sigma$, $\bar{w} = \tau - \sigma$, the UR limit is 
\be{URlim}
\s \to \s, \t \to \e \t, \, \e \to 0.
\ee
This is a limit where the speed of light in this 2d field theory goes to zero and hence the theory is a non-Lorentzian theory. This closing of lightcones is a peculiar feature and these theories have been dubbed Carrollian theory. It is easy to check that the limit \refb{URlim} on \refb{vir-cyl} naturally generates the linear combinations of \refb{cft2bms-ur} and results in the expressions of the BMS generators we presented earlier in \refb{bms-cyl}. Using this map \refb{URlim}, the map between the eigenvalues of $\L_0, \bcL_0$ and $L_0, M_0$ simply becomes
\be{}
h_L= h-\h, \quad h_M=\e(h+\h).
\ee

We mentioned a peculiar duality in $d=2$ between the ultra-relativistic or $c\to0$ and the non-relativistic or $c\to\infty$ theories in the introduction. It is instructive to check that the limit diametrically opposite to \refb{URlim} \cite{Bagchi:2009my}:
\be{NRlim}
\s \to \e \s, \t \to  \t, \, \e \to 0
\ee
that send the speed of light to infinity instead of zero results in the linear combinations \cite{Bagchi:2009pe}:
\be{cft2bms-nr}
L_{n}=\mathcal{L}_n+\mathcal{\bar{L}}_{n}, \qquad  M_n=\epsilon(\mathcal{L}_n-\mathcal{\bar{L}}_{n}),
\ee
which again yield the BMS from two copies of the Virasoro. The generators that arise from contracting \refb{vir-cyl} using the non-relativistic (NR) limit \refb{NRlim} are
\be{}
L_n=ie^{in\t}(\partial_{\t}+in\sigma\partial_{\s}), \qquad M_n=ie^{in\t}\partial_{\s}.
\ee
which is identical to \refb{bms-cyl} with an interchange of $\t \leftrightarrow \s$. The limit from 2d CFT to BMS$_3$ gives us a vital tool for cross-checking answers obtained by just using symmetries of BMS. Many of the answers, e.g. the expressions of the BMS EM tensors, can be easily arrived at when we look at the corresponding CFT expressions and carefully implement the limit. 

\subsection{From Carroll to BMS}

Generic null manifolds are endowed with a Carrollian structure that replaces the Riemannian structure due to the lack of a non-degenerate metric. The basic geometric quantities that locally describe a Carrollian manifold are a no-where vanishing vector field $\zeta$ and a degenerate metric $\h$. As we remarked earlier in our introduction, BMS structures arise when we look at the conformal version of Carrollian symmetry. In this case, we can look at the conformal isometries of Carrollian manifolds we have just defined \cite{Duval:2014lpa,Duval:2014uoa,Duval:2014uva}. The conformal structures can be defined on these manifolds by
\be{car}
	\mathfrak{L}_{\xi}\zeta=\lambda_1 \zeta , \qquad \mathfrak{L}_{\xi}\h=\lambda_2\h ,
\ee
where $\mathfrak{L}$ is the Lie derivative of these fields under a co-ordinate transformation by $\xi$. 
The standard flat Carroll structure, arising from a generalisation of usual Minkowskian spacetime, is given by a topology $\mathbb{R} \times \mathbb{R}^d$. In a coordinate chart $(u, x^i)$ with $i=1, \ldots d$
\be{flatc}
\zeta = \frac{\p}{\p u}, \quad \h = \delta_{ij} dx^i dx^j.
\ee 
Solving the above equations \refb{car} in the case of the flat Carroll structure \refb{flatc}, one can find an expression for the vector field $\xi$. The individual independent components close to what is called the conformal Carroll algebra. When space and time dialate in the same way (in the language above this amounts to $\lambda_2/\lambda_1=-2$), as is expected when one is computing the asymptotic symmetries of a relativistic spacetime, the symmetry algebra can be shown to be isomorphic to the BMS algebra. The process of UR contractions mentioned above also lead us to the same vector fields. Thus the limiting analysis is a rather nice way of understanding such manifolds and the physics on these manifolds, starting out from relativistic physics and usual Riemannian manifolds.


\section{Warming up: delta functions, Ward identities and correlators}
Before we delve into the details of the calculation of the Weyl anomaly for BMSFTs, in this section, we touch upon some of the ingredients we would require for our calculations. We begin by presenting the OPE of the BMS energy momentum tensors. A delta-function identity would be required for the Weyl anomaly calculation. We would present such an identity for Carrollian manifolds. As a check of our proposed identity, we would then reproduce the Ward identities and the known form of the stress-energy two point function of the 2d BMSFT. We will first present analysis on the null plane before indicating how the same works for the null cylinder. 

\subsection{BMSFT on a null plane}

We begin with a 2d BMSFT on a null plane which is a $(1+1)$ dimensional flat Carrollian background with topology ${\rm I\!R}_{null}\times {\rm I\!R}$. Let $t$ and $x$ be the coordinates along ${\rm I\!R}_{null}$ and ${\rm I\!R}$ respectively. We propose a representation of the delta function on a null plane:
\be{deltafn-plane}
\partial_{t_1}\partial_{x_1}G(t_1,x_1;t_2,x_2) = 2\pi \delta^{(2)}(t_{12},x_{12});  \quad G(t_1,x_1;t_2,x_2) = \log (t_{12} x_{12}) 
\ee
where $x_{ij}= x_i- x_j$. Now for our other basic ingredient, the BMS stress-tensor OPE. On the null plane this reads
\bea{TTbms}
	T_1(t_1,x_1)T_1(t_2,x_2) &\sim& \frac{c_L}{2x_{12}^4}-\frac{2t_{12} c_M}{x_{12}^5} + \frac{2T'_1}{x_{12}^2} -\frac{4t_{12} T'_2}{x_{12}^3} +\frac{\partial_{x_2}T'_1}{x_{12}}-\frac{t_{12} (\partial_{t_2} T'_1+ \p_{x_2} T'_2)}{2x_{12}^2} + \text{reg}, \label{T1T1ope-plane} \nonumber\\
	T_1(t_1,x_1)T_2(t_2,x_2) &\sim& \frac{c_M}{2x_{12}^4} + \frac{2T'_2}{x_{12}^2}+\frac{\partial_{x'}T'_2}{x_{12}} + \text{reg},  \\
	T_2(t_1,x_1)T_2(t_2,x_2) &\sim & \text{reg}.  \nonumber
\end{eqnarray}
Here we use the shorthand $T'\equiv T(x_2, t_2)$. The above can be derived intrinsically from the symmetries of the algebra or from the limit of the relativistic $TT$ OPE. From this, it is easy to derive the two-point correlation functions of the BMS EM tensor. These are given by
\be{TT2pt}
\<T_1(t_1,x_1)T_1(t_2,x_2)\> = \frac{c_L}{2x_{12}^4}-\frac{2t_{12} c_M}{x_{12}^5}, \quad \<T_1(t_1,x_1)T_2(t_2,x_2)\> = \frac{c_M}{2x_{12}^4},
\ee
while $\<T_2T_2\>$ vanishes. Again, these can be derived entirely from the algebra of the generators \refb{BMSalgebra-ce}. We shall now rederive these results from using the delta-function identity we have proposed. 

\subsection*{Ward identities and stress-energy correlators}\label{sec:Wid-plane}
In this part, we shall be adopting techniques outlined in \cite{Bagchi:2015wna}. The calculations there were performed on the null cylinder. Our formulation in this subsection would be on the plane, before moving onto the cylinder in the next subsection for completeness. We also have a non-zero $c_L$. So the results are somewhat more general. We begin by considering a deformation to a 2d free BMSFT on a Carrollian plane, described by an action $S_0$, by sources $\mu_L$ and $\mu_M$ for the stress tensor components\,:
\begin{equation}
	S_{\mu} = S_0 -\int dt dx \big[\mu_L(t,x) T_1(t,x) + \mu_M(t,x) T_2(t,x) \big] \ ,
\end{equation}
where we localize the sources at some point $(t',x')$ as
\begin{equation}
	\mu_L(t,x) = \epsilon_L\delta^{(2)}(t-t',x-x'), \qquad \mu_M(t,x) = \epsilon_M\delta^{(2)}(t-t',x-x')\ . \label{mu-sources-plane}
\end{equation}
Here $\epsilon_L$ and $\epsilon_M$ are small expansion parameters. The expectation value of $T_1$ in the $\mu$-deformed theory gives the $2$-point correlators of $T_1$ and $T_2$ in the free theory as
\bea{T1expansion-plane}
\langle T_1(t, x) \rangle_{\mu} &=& \langle T_1 (t, x) \exp{\int dt''dx''\left(\mu_L(t'',x'') T_1(t'', x'') + \mu_M(t'', x'') T_2(t'', x'')\right)}\rangle_{0} \nonumber\\
&=& \langle T_1(t, x)\rangle_0 + \epsilon_L\langle T_1(t, x) T_1(t', x')\rangle_0 + \epsilon_M\langle T_1 (t, x) T_2(t', x')\rangle_0 + O(\epsilon^2).
\eea
Similarly for $T_2$:
\be{T2expansion-plane}
\langle T_2(t, x) \rangle_{\mu} = \langle T_2(t, x)\rangle_0 + \epsilon_L\langle T_2(t, x) T_1(t', x')\rangle_0 + \epsilon_M\langle T_2 (t, x) T_2(t', x')\rangle_0 + O(\epsilon^2).
\ee
Now we define $\mathcal{M} = \langle T_2\rangle_{\mu}$, $\mathcal{N} = \langle T_1\rangle_{\mu}$ and expand in $\epsilon_{L/M}$ as
\begin{equation}
	\mathcal{M} = \mathcal{M}^{(0)} + \mathcal{M}^{(1)} + \cdots, \qquad \mathcal{N} = \mathcal{N}^{(0)} + \mathcal{N}^{(1)} + \cdots, \label{MN-expansion-plane}
\end{equation}
where $\mathcal{M}^{(n)} \sim O(\epsilon^n)$ and so on. Then comparing with \eqref{T1expansion-plane} and \refb{T2expansion-plane}, we have
\begin{eqnarray}
	&& \mathcal{M}^{(0)} = \langle T_2\rangle_0, \qquad \mathcal{M}^{(1)} = \epsilon_L\langle T_2T'_1\rangle_0 +\epsilon_M\langle T_2T'_2\rangle_0, \nonumber \\
	&& \mathcal{N}^{(0)} = \langle T_1\rangle_0 , \qquad \mathcal{N}^{(1)} = \epsilon_L\langle T_1T'_1\rangle_0 +\epsilon_M\langle T_1T'_2\rangle_0 . \label{MN-correlators}
\end{eqnarray}
Here we have reverted to the shorthand $T_1\equiv T_1(t,x)$, $T'_1\equiv T_1(t',x')$, and so on. 
The classical conservation equations in the free theory can be written as
\begin{eqnarray}
	\partial_t T_2 = 0 \, \rightarrow \, \partial_t \mathcal{M}^{(0)} = 0, \quad \partial_t T_1 = \partial_x T_2 \, \rightarrow \, \partial_t \mathcal{N}^{(0)} = \partial_x \mathcal{M}^{(0)}.
\end{eqnarray}
To derive Ward identities we compute derivatives of $\langle T_1\rangle_{\mu}$ and $\langle T_2\rangle_{\mu}$ as follows. Let us first consider the time derivative of $\langle T_2\rangle_{\mu}$\,:
\begin{eqnarray}
	\partial_t \langle T_2\rangle_{\mu} = \langle (-\partial_t S_{\mu})T_2 + \partial_t T_2 \rangle_{\mu} = \Big\langle \int dt'dx' \Big(\mu'_L \partial_t (T_2T'_1) + \mu'_M \partial_t (T'_2 T_2) \Big) \Big\rangle_{\mu} ,
\end{eqnarray}
where we have used $\partial_t S_0 = 0$ and a conservation equation $\partial_t T_2 = 0$. We use the OPEs \refb{TTbms} to simplify $\partial_t \langle T_2\rangle_{\mu}$ further as
\begin{eqnarray}
	\partial_t \langle T_2\rangle_{\mu} &=& \Big\langle \int dt'dx'\mu'_L\partial_t \Big(\frac{c_M}{2(x-x')^4} + \frac{2T'_2}{(x-x')^2}+\frac{\partial_{x'}T'_2}{(x-x')} + reg \Big) \Big\rangle_{\mu} \nonumber \\
	&=& \Big\langle \int dt'dx' \mu'_L \Big[ \frac{c_M}{12}\partial_t\partial^2_x\partial'_x\Big(\frac{1}{\Delta x}\Big) -2T'_2\partial_t\partial_x\Big(\frac{1}{\Delta x}\Big) +\partial_{x'}T'_2\partial_t\Big(\frac{1}{\Delta x}\Big) \Big] \Big\rangle_{\mu}
\end{eqnarray}
where $\Delta x = x-x'$ and $\Delta t = t-t'$. Using the delta function \eqref{deltafn-plane} as $$\partial_t \Big(\frac{1}{\Delta x} \Big) = 2\pi \delta^{(2)}(\Delta t, \Delta x)\equiv 2\pi \delta^{(2)},$$ we get
\begin{eqnarray}
	\partial_t \langle T_2\rangle_{\mu} &=& \Big\langle 2\pi \int dt'dx' \mu'_L \Big[ \frac{c_M}{12}\partial^2_x\partial'_x\delta^{(2)} -2T'_2\partial_x\delta^{(2)}+\partial_{x'}T'_2\delta^{(2)} \Big] \Big\rangle_{\mu} \nonumber \\
	&=& \Big\langle 2\pi \Big( -\frac{c_M}{12}\partial^3_x\mu_L -2\partial_x(T_2\mu_L) +\mu_L\partial_{x}T_2 \Big) \Big\rangle_{\mu}, \nonumber \\
 \Rightarrow -\frac{1}{2\pi}\partial_t\mathcal{M} &=& \frac{c_M}{12}\partial^3_x\mu_L +2\mathcal{M}\partial_x\mu_L +\mu_L\partial_{x}\mathcal{M} . \label{Wid-plane-M}
\end{eqnarray}
Thus we obtain the first Ward identity for the stress tensor component $\mathcal{M} = \partial_t \langle T_2\rangle_{\mu}$. Expanding $\mathcal{M}$ as in \eqref{MN-expansion-plane} and using \eqref{mu-sources-plane}, the leading term gives a conservation equation $\partial_t\mathcal{M}^{(0)} = 0$ and $O(\epsilon)$ terms give a Ward identity for $\mathcal{M}^{(1)}$ as
\begin{equation}
	-\frac{1}{2\pi}\partial_t \mathcal{M}^{(1)} = \epsilon_L\Big[\frac{c_M}{12}\partial^3_x\delta^{(2)} +2\mathcal{M}^{(0)}\partial_x\delta^{(2)} +\delta^{(2)}\partial_{x}\mathcal{M}^{(0)} \Big] . \label{Wid-plane-M1}
\end{equation}
Now let us consider the time derivative of $\langle T_1\rangle_{\mu}$\,:
\be{}
\partial_t \langle T_1\rangle_{\mu} = \Big\langle \int dt'dx' \Big(\mu'_L \partial_t (T_1T'_1) + \mu'_M \partial_t (T_1T'_2) \Big) + \partial_t T_1 \Big\rangle_{\mu}
\ee 
Using the form of the OPEs, the delta function \eqref{deltafn-plane}, and the conservation equation $\partial_tT_1 = \partial_xT_2$ (details in Appendix \ref{sec:Wid-plane-detials}), the above gives rise to our second Ward identity
\bea{}	
 \frac{1}{2\pi}(\partial_x\mathcal{M} -\partial_t\mathcal{N}) = \frac{c_L}{12}\partial^3_x\mu_L +2\mathcal{N}\partial_x\mu_L + \mu_L\partial_x\mathcal{N} +\frac{c_M}{12}\partial^3_x\mu_M +2\mathcal{M}\partial_x\mu_M+\mu_M\partial_x\mathcal{M}. \nonumber\\\label{Wid-plane-N}
\eea
Expanding $\mathcal{N}$ and $\mathcal{M}$ as in \eqref{MN-expansion-plane} and using \eqref{mu-sources-plane}, the leading term gives the conservation equation $\partial_t\mathcal{N}^{(0)} = \partial_x\mathcal{M}^{(0)}$ and $O(\epsilon)$ terms give a Ward identity for $\mathcal{N}^{(1)}$ as
\begin{eqnarray}
	-\frac{1}{2\pi}(\partial_t \mathcal{N}^{(1)} - \partial_x\mathcal{M}^{(1)}) = && \epsilon_L\Big[\frac{c_L}{12}\partial^3_x\delta^{(2)} +2\mathcal{N}^{(0)}\partial_x\delta^{(2)} +\delta^{(2)}\partial_{x}\mathcal{N}^{(0)} \Big] \nonumber \\ &&+ \epsilon_M\Big[\frac{c_M}{12}\partial^3_x\delta^{(2)} +2\mathcal{M}^{(0)}\partial_x\delta^{(2)} +\delta^{(2)}\partial_{x}\mathcal{M}^{(0)} \Big]. \label{Wid-plane-N1}
\end{eqnarray}
Now using the values $\mathcal{M}^{(0)} = 0 = \mathcal{N}^{(0)}$ on a plane and the delta function \eqref{deltafn-plane}, we integrate the above Ward identities \eqref{Wid-plane-M1} and \eqref{Wid-plane-N1} to get
\begin{eqnarray}
	&& \mathcal{M}^{(1)} = \epsilon_L\frac{c_M}{2\Delta x^4}, \quad \mathcal{N}^{(1)} = \epsilon_L\Big(\frac{c_L}{2\Delta x^4} - \frac{2c_M\Delta t}{\Delta x^5} \Big) + \epsilon_M\frac{c_M}{2\Delta x^4}\ .
\end{eqnarray}
Comparing these with \eqref{MN-correlators}, we get the desired $\langle T_i T_j\rangle$ correlators as
\begin{equation}
	\langle T_1(t,x) T_1(t', x')\rangle = \frac{c_L}{2\Delta x^4} - \frac{2c_M\Delta t}{\Delta x^5}, \quad \langle T_1(t,x) T_2(t', x')\rangle = \frac{c_M}{2\Delta x^4}, \quad \langle T_2(t,x) T_2(t', x')\rangle = 0.
\end{equation}
These are of course in keeping with our initial observations \refb{TTbms} and hence a robust check of the validity of our proposed delta function 
identity. The method above can be generalised to compute arbitrary point correlation functions for BMS stress tensors on the null plane. 

\subsection{BMSFT on a null cylinder}
We now present the analogous results of the analysis in the previous subsection for the case of 2d BMSFTs on the null cylinder.  The null cylinder is a flat Carrollian manifold with topology ${\rm I\!R}_{null}\times S^1$. Let $\tau$ be the null time coordinate along ${\rm I\!R}_{null}$ and $\sigma \sim \sigma + 2\pi$ be the spatial coordinate parametrizing $S^1$. Similar to the case of BMSFT on a null plane, we need a representation of the delta function. A delta function on a null cylinder, introduced in \cite{Bagchi:2015wna}, is
\begin{eqnarray}
	&& \partial_{\tau}\partial_{\sigma}G(\tau-\tau',\sigma-\sigma')=2\pi \delta^{(2)}(\tau-\tau',\sigma-\sigma') ; \nonumber \\
	&& G(\tau-\tau',\sigma-\sigma')=\log\Big[(\tau-\tau')\sin\Big(\frac{\sigma-\sigma'}{2}\Big)\Big] , \label{deltafn-cylinder}
\end{eqnarray}
with the normalization $\int d\tau d\sigma \delta^{(2)}(\tau,\sigma) = 1$. This delta function was already justified in \cite{Bagchi:2015wna} by reproducing correlators of stress tensor components in a BMSFT on a null cylinder, dual to Einstein gravity with $c_L = 0$. However for completeness and for generic theories with $c_L \neq 0$, we briefly discuss the Ward identities and $2$-point correlators of stress tensor components in appendix \ref{sec:Wid-cylinder}. The important results are listed below. 

\paragraph{TT OPE on cylinder:}
\begin{subequations}
\label{TTope-cylinder}
	\begin{eqnarray}
		T_1^{(1)}T_1^{(2)} &\sim& \frac{c_L}{2s_{12}^4} -\frac{c_L}{12s_{12}^2} -\frac{2c_M\tau_{12}c_{12}}{2s_{12}^4} +\frac{2c_M\tau_{12}c_{12}}{24s_{12}^2} +\frac{2T_1^{(2)}}{s_{12}^2}-\frac{2T_2^{(2)}\tau_{12}c_{12}}{s_{12}^2} \nonumber \\
		&& +\frac{\partial_{\sigma_2}T_1^{(2)}}{s_{12}} -\frac{(\partial_{\tau_2}T_1^{(2)}+\partial_{\sigma_2}T_2^{(2)})\tau_{12}c_{12}}{4s_{12}} + {\text{reg}},\\
		T_1^{(1)}T_2^{(2)} &\sim & \frac{c_M}{2s_{12}^4}-\frac{c_M}{12s_{12}^2}+\frac{2T_2^{(2)}}{s_{12}^2} +\frac{(\partial_{\tau_2}T_1^{(2)}+\partial_{\sigma_2}T_2^{(2)})}{2s_{12}} + {\text{reg}}, \\
		T_2^{(1)}T_2^{(2)} &\sim& \text{reg},
	\end{eqnarray}
\end{subequations}
where we have defined $T_a^{(i)} \equiv T_a(\tau_i,\sigma_i)$ for $a=1,2$, and $s_{ij}\equiv 2\sin(\frac{\sigma_i - \sigma_j}{2})$, $c_{ij}\equiv \cot(\frac{\sigma_i - \sigma_j}{2})$, $\tau_{ij} = \tau_i - \tau_j$. Here `reg' denotes regular terms as $(\tau_1,\sigma_1)\rightarrow (\tau_2,\sigma_2)$.\\

\paragraph{Ward identities:} 
\bea{}
-\frac{1}{2\pi}\partial_{\tau}\mathcal{M} &=& \frac{c_M}{12}\partial^3_{\sigma}\mu_L + 2\mathcal{M}\partial_{\sigma}\mu_L + \mu_L\partial_{\sigma}\mathcal{M}. \\
-\frac{1}{2\pi}(\partial_{\tau}\mathcal{N} - \partial_{\sigma}\mathcal{M})&=& \frac{c_L}{12}\partial^3_{\sigma}\mu_L + 2\mathcal{N}\partial_{\sigma}\mu_L + \mu_L\partial_{\sigma}\mathcal{N}  +\frac{c_M}{12}\partial^3_{\sigma}\mu_M + 2\mathcal{M}\partial_{\sigma}\mu_M + \mu_M\partial_{\sigma}\mathcal{M}. \nonumber
\eea

\medskip

\noindent As was stated before, these results generalise the analysis of \cite{Bagchi:2015wna} to non-zero values of $c_L$. The formulae would be useful when we derive the BMS version of the Weyl anomaly for the BMSFTs on the null cylinder. 

\section{BMS-Weyl symmetry and trace anomaly}\label{sec:BMSWeyl}
In this section, we study aspects of BMS-Weyl symmetry and trace anomaly in BMSFTs in $(1+1)$ dimensions. These theories live on Carrollian background spacetimes and we will mostly focus on field theories coupled to flat Carrollian backgrounds, using zweibein formulation (see appendix \ref{sec:zweibeinformulation} for some details).

\subsection{Weyl invariance and stress tensor}
In relativistic CFTs, a consequence of conformal invariance is vanishing of the trace of the stress tensor classically. Analogously in BMSFTs, the BMS-Weyl symmetry also leads to vanishing of the trace of the stress tensor. To see this, consider a 2d BMSFT described by an action $S[\phi,e^a_{\alpha}]$ for some field $\phi(\tau,\sigma)$ on a Carrollian background described by zweibeins $e^a_{\alpha}(\tau,\sigma)$. Here $\tau$ and $\sigma$ are null time and spatial coordinates respectively. We define the stress tensor in this theory as a variation of the action with respect to zweibeins as
\begin{equation}\label{stresstensor-defn}
	T^{\alpha}_{\ \beta} \equiv \frac{e^{\alpha}_a}{2e} \frac{\delta S}{\delta e^{\beta}_a}\ ,
\end{equation}
where $e = \det(e^a_{\alpha})$ is the determinant. In terms of zweibeins, the BMS-Weyl transformation is given by 
\begin{equation}\label{BMS-Weyl}
	e^a_{\alpha} \rightarrow e^{\Omega(\tau,\sigma)}e^a_{\alpha}\ , \qquad e^{\alpha}_a \rightarrow e^{-\Omega(\tau,\sigma)}e^{\alpha}_a\ ; \qquad e \rightarrow e^{2\Omega(\tau,\sigma)}e\ .
\end{equation}
The above definition is equivalent to the one in the existing literature \cite{Ciambelli:2018wre,Ciambelli:2018xat,Ciambelli:2019lap, Gupta:2020dtl}. (See Appendix \ref{sec:zweibeinformulation} for some further discussion on this.) 

\medskip

\noindent For small $\Omega(\tau,\sigma)$, we have an infinitesimal BMS-Weyl transformation $\delta e^{\alpha}_a = -\Omega e^{\alpha}_a$ under which the action $S[\phi,e^a_{\alpha}]$ changes as
\begin{eqnarray}
	\delta S &=& \int d^2\sigma 2\,e\, e^a_{\alpha}T^{\alpha}_{\ \beta}\delta e^{\beta}_a = -2\int d^2\sigma\,e\,\Omega\, T^{\alpha}_{\ \alpha}\ . \label{deltaS-BMS-Weyl}
\end{eqnarray}
Thus we see that invariance of the theory under BMS-Weyl transformation, \ie\ $\delta S = 0$ leads to vanishing of the trace of the stress tensor. 

\subsection{Avoiding the Diff Anomaly}
We saw briefly in Sec.~2 that our discussions of the Weyl anomaly can be complicated by the presence of a diffeomorphism anomaly in 2d CFTs which is reflected in the difference between the central terms $c-\bar{c}$. This is reflective of a violation of parity in the 2d theory. In our discussions of 2d BMSFTs, we wish to also avoid the diffeomorphism anomaly in our present considerations. In these 2d BMSFTs, there is no a priori notion of left and right movers and hence perhaps it is somewhat unclear what one needs to do to avoid the gravitational anomaly. Here we give an explanation of this from the point of the proposed bulk dual, as well as some purely field theoretic arguments. 

\medskip

\noindent
2d BMSFTs are putative duals to gravitational theories in 3d asymptotically flat spacetimes. When one considers 3d Einstein gravity, the central charges of the dual BMSFT are given by:
\be{}
c_L = 0, \quad c_M= \frac{3}{G}.
\ee
We note here that Einstein gravity is a parity invariant theory. In order to get a theory with a non-zero $c_L$, one needs to add a gravitational Chern-Simons (GCS) term to the Einstein-Hilbert action. The resulting theory, called Topologically Massive Gravity \cite{Deser:1982vy}, thus has an action
\bea{}
S_{\text{TMG}} &=& S_{\text{EH}} + S_{\text{GCS}} + S_{\text{bdy}} \\ &=& \int d^3x \sqrt{-g} \left\{\mathcal{R} + \frac{1}{2\mu} \varepsilon^{\lambda\mu\nu}\Gamma^\rho_{\, \lambda \s} \left( \p_\mu \Gamma^\s_{\, \rho \nu} + \frac{2}{3} \Gamma^\s_{\, \mu \t} \Gamma^\t_{\, \nu \rho}\right) \right\}+ S_{\text{bdy}} \nonumber
\eea
where $S_{\text{bdy}}$ is a boundary term put in to make the variational problem well defined \cite{Bagchi:2013lma, Detournay:2014fva}. 
An asymptotic symmetry analysis of its boundary following with appropriate asymptotically flat boundary conditions \cite{Bagchi:2012yk} yields that the ASG is again given by the BMS$_3$, where now both central terms are switched on and are 
\be{}
c_L = \frac{3}{\mu G}, \quad c_M= \frac{3}{G}.
\ee
The GCS term clearly violates parity and induces a handedness in the theory. Hence in terms of the dual BMSFT, the violation of parity is encoded in switching on a non-zero $c_L$. This is also clear from our discussions of the ultra-relativistic limit, where 
\be{}
c_L = c - \bar{c}, \quad c_M = \e (c + \bar{c}). 
\ee
Finally, just from the point of view of the boundary theory, without evoking a dual bulk or the existence of a limit, there is a clear hint of this condition. Let us envision a situation where 
\be{}
c_L = 24k, \quad c_M=0
\ee
where $k$ some constant. By an analysis of null vectors, it can be shown \cite{Bagchi:2009pe, Bagchi:2012yk} that in this case, the BMS$_3$ algebra reduces to a single copy of a Virasoro algebra. This is also known as the chiral limit of BMS$_3$ in literature, and clearly this is a limit where the resulting field theory is purely chiral. This limit, from the point of view of the dual bulk, is where one takes the double scaling limit
\be{}
G \to \infty, \mu \to 0, \quad \text{keeping fixed} \, \, \mu G = \frac{1}{8k}.
\ee
The gravitational theory then reduces to just the GCS term and this is what is called Chern-Simons gravity. The conjecture of CS gravity with asymptotically flat boundary conditions being dual a chiral half of a 2d CFT is known as the Flatspace Chiral Gravity conjecture \cite{Bagchi:2012yk}. For some further work on this, the reader is referred to \cite{Bagchi:2013hja, Grumiller:2015xaa, Bagchi:2018ryy}. 

\medskip

\noindent The upshot of all this discussion is the fact that $c_L\neq0$ theories are naturally parity violating and would lead to diffeomorphism anomalies in the 2d BMSFTs. This is a complication that we will look to avoid in the current paper and will come back to in future work. 

\subsection{BMS Trace Anomaly}\label{sec:BMSanomaly}
For relativistic 2d CFTs coupled to non-trivial background, the trace of the stress tensor, which vanishes classically, suffers from a quantum anomaly. This trace anomaly arises because the underlying Virasoro symmetry algebra admits a central extension. We are now interested in quantum field theories that are BMS invariant. The BMS$_3$ algebra \eqref{BMSalgebra-ce}, like the Virasoro, is centrally extended. The expectation is that these central extensions in BMS$_3$ algebra would lead to a trace anomaly in 2d BMSFTs. In what follows, we will obtain expression for this BMS trace or Weyl anomaly . 

\medskip

\noindent As we saw in Sec.~\ref{sec:2DCFT}, in relativistic CFTs, the trace anomaly can be computed by considering an infinitesimal Weyl transformation of the flat background, \ie\ $\delta g_{\mu\nu}=2\Omega \delta_{\mu\nu}$ and expanding the trace of the stress tensor in powers of $\Omega$. We now proceed to obtain the trace anomaly in 2d BMSFTs by doing an analogous computation. To do so, we consider a 2d BMSFT on a flat Carrollian background described by zweibeins $$e^a_{\mu} = \delta^a_{\mu} \quad \text{and} \, e^{\mu}_a = \delta^{\mu}_a,$$ where $a=1,2$ and $\mu=\tau,\sigma$ are tangent space and spacetime indices respectively. (See appendix \ref{sec:zweibeinformulation} for more details on Carrollian geometry.) Consider that we are performing an infinitesimal Weyl transformation of the flat Carrollian background:
$$e^a_{\alpha}(\tau,\sigma) =e^{\Omega(\tau,\sigma)}\delta^a_{\alpha} \implies \delta e^a_{\alpha}(\tau,\sigma) = \Omega(\tau,\sigma)\delta^a_{\alpha}$$ for $\Omega(\tau,\sigma)$ small. Then in the quantum theory on a background described by $e^a_{\alpha}(\tau,\sigma)$, the expectation value of the trace of the stress tensor can be expanded in powers of $\Omega$ as
\begin{equation}
	\langle T^{\alpha}_{\ \alpha}(\tau,\sigma)\rangle_{e} = \langle T^{\alpha}_{\ \alpha}(\tau,\sigma)\rangle_{\delta} +\delta\langle T^{\alpha}_{\ \alpha}(\tau,\sigma)\rangle_{\delta} + \mathcal{O}(\Omega^2) ,
\end{equation}
where $\delta\langle T^{\alpha}_{\ \alpha}(\sigma)\rangle_{\delta}\sim \mathcal{O}(\Omega)$ and the subscript $e$ denotes that the quantity is evaluated in the theory on a background described by $e^a_{\alpha}$. Translational invariance implies that $\langle T^{\alpha}_{\ \alpha}(\tau,\sigma)\rangle_{\delta} = 0$ for a flat Carrollian background. The expansion of the trace becomes
\begin{equation}\label{traceexpansion}
	\langle T^{\alpha}_{\ \alpha}(\tau,\sigma)\rangle_{e} = \delta\langle T^{\alpha}_{\ \alpha}(\tau,\sigma)\rangle_{\delta} + \mathcal{O}(\Omega^2) . 
\end{equation}
Using the definition of the expectation value $$\langle T^{\alpha}_{\ \alpha}(\tau,\sigma)\rangle_{e} = \frac{1}{Z}\int \mathcal{D}\phi e^{- S[\phi,e^a_{\alpha}]}T^{\alpha}_{\ \alpha}(\tau,\sigma)$$ and \eqref{deltaS-BMS-Weyl}, we can compute the variation $\delta\langle T^{\alpha}_{\ \alpha}(\tau,\sigma)\rangle_{\delta}$ under an infinitesimal Weyl transformation $\delta e^a_{\alpha}(\tau,\sigma) = \Omega(\tau,\sigma)\delta^a_{\alpha}$ as
\begin{eqnarray}
	\delta \langle T^{\alpha}_{\ \alpha}(\tau,\sigma)\rangle_{\delta} &=& \frac{1}{Z}\int \mathcal{D}\phi e^{- S[\phi,\delta^a_{\alpha}]}(-\delta S) T^{\alpha}_{\ \alpha}(\tau,\sigma) \nonumber \\
	&=& \frac{1}{Z}\int \mathcal{D}\phi e^{-S[\phi,\delta^a_{\alpha}]}T^{\alpha}_{\ \alpha}(\tau,\sigma) \int d^2\sigma'\,(-2)e(\tau',\sigma')\,e^a_{\beta}(\tau',\sigma')T^{\beta}_{\ \rho}(\tau',\sigma')\delta e^{\rho}_a(\tau',\sigma')  \nonumber \\
	&=& \frac{1}{Z}\int \mathcal{D}\phi e^{-S[\phi,\delta^a_{\alpha}]} \int d^2\sigma' 2\, T^{\alpha}_{\ \alpha}(\tau,\sigma) T^{\beta}_{\ \beta}(\tau',\sigma')\Omega(\tau',\sigma') , \label{deltatrace}
\end{eqnarray}
where we have used $e = 1$ for $e^a_{\alpha}(\tau,\sigma) = \delta^a_{\alpha}$.

\medskip

\noindent We know that  the trace of the stress tensor vanishes classically. However in quantum theory, the $T^\alpha_{\ \alpha} T'^\beta_{\ \beta}$ OPE is non-vanishing and leads to contact terms which contribute to the anomaly, as can be seen from \eqref{deltatrace}. Following a computation as in relativistic CFTs, we now obtain the $T^\alpha_{\ \alpha} T'^\beta_{\ \beta}$ OPE using the OPEs of the stress tensor components and the stress tensor conservation. The $\beta = \tau,\sigma$ components of the stress tensor conservation equation $\partial_\alpha T^{\alpha}_{\ \beta}=0$ are
\begin{eqnarray}
 \partial_{\tau}T^{\tau}_{\ \tau} + \partial_{\sigma}T^{\sigma}_{\ \tau} = 0, \, \partial_{\tau}T^{\tau}_{\ \sigma} + \partial_{\sigma}T^{\sigma}_{\ \sigma} = 0. \label{conservationsigma}
\end{eqnarray}
From sec.~\ref{sec:BMStheories}, we have $T^{\alpha}_{\ \alpha} = T^{\tau}_{\ \tau} + T^{\sigma}_{\ \sigma} = 0$ classically and $T^{(cl)}_1 = T^{\tau}_{\ \sigma}$, $T^{(cl)}_2 = T^{\tau}_{\ \tau} = -T^{\sigma}_{\ \sigma}$. However, in the quantum theory, anticipating a non-vanishing $T^{\alpha}_{\ \alpha}$, we define
\begin{equation}
	T^{\tau}_{\ \sigma} = T_1 , \qquad T^{\tau}_{\ \tau} = T_2 + \frac{T^{\alpha}_{\ \alpha}}{2} , \qquad T^{\sigma}_{\ \sigma} = - T_2 + \frac{T^{\alpha}_{\ \alpha}}{2}.
\end{equation}
Then the second conservation equation \eqref{conservationsigma} can be written as
\begin{equation}
	\partial_{\tau} T_1 - \partial_{\sigma}T_2 + \frac{1}{2}\partial_{\sigma}T^{\alpha}_{\ \alpha} = 0 ,
\end{equation}
which leads to a differential equation for $T^{\alpha}_{\ \alpha}(\tau,\sigma)T^{\beta}_{\ \beta}(\tau',\sigma')$ as
\begin{eqnarray}\label{traceeqn}
	\partial_{\sigma}\partial_{\sigma'}T^{\alpha}_{\ \alpha} T'^{\beta}_{\ \beta} = 4\big( \partial_{\sigma}\partial_{\sigma'}T_2T_2' -\partial_{\sigma}\partial_{\tau'}T_2T_1' -\partial_{\tau}\partial_{\sigma'}T_1T_2' +\partial_{\tau}\partial_{\tau'}T_1T_1' \big) .
\end{eqnarray}
Here we have used the shorthand $T_1 \equiv T_1(\tau,\sigma)$, $T'_1 \equiv T_1(\tau',\sigma')$ and so on. The analysis leading to the differential equation \eqref{traceeqn} for the $T^{\alpha}_{\ \alpha} T'^{\beta}_{\ \beta}$ OPE is generic for a BMS$_3$ invariant field theory on any flat Carrollian background. In subsequent sections, we focus on two particular flat Carrollian backgrounds\,: a degenerate plane and a degenerate cylinder.

\subsection{Anomaly for null plane}
We begin on the null plane, where the null time coordinate is given by $t$ and the unwrapped spatial coordinate by $x$. The BMS generators are given by \refb{bms-gen-pl}. The relevant delta-function identity and the EM-tensor OPEs we would be using are given by \refb{deltafn-plane} and \refb{TTbms} respectively. We will be turning off $c_L$ in order to avoid complications arising out of diffeomorphism anomalies and the possible non-conservation of the stress tensor, as discussed earlier. Armed with this information, we now write the differential equation \eqref{traceeqn} for the $T^{\alpha}_{\ \alpha}T'^{\beta}_{\ \beta}$ OPE in $(t,x)$ coordinates:
\begin{eqnarray}
	\partial_{x}\partial_{x'} T^{\alpha}_{\ \alpha}(t,x)T^{\beta}_{\ \beta}(t',x') &=& 4\big(\partial_{t}\partial_{t'}T_1(t,x)T_1(t',x') - \partial_{t}\partial_{x'}T_1(t,x)T_2(t',x') \nonumber \\
	&& - \partial_{x}\partial_{t'}T_1(t,x)T_2(t',x') + \partial_{x}\partial_{x'}T_2(t,x)T_2(t',x') \big). \label{traceope-plane-diffeqn}
\end{eqnarray}
Substituting the central terms in the OPEs \eqref{T1T1ope-plane} in \eqref{traceope-plane-diffeqn} and simplifying, we get
\begin{eqnarray}
	\partial_{x}\partial_{x'} T^{\alpha}_{\ \alpha}(t,x)T^{\beta}_{\ \beta}(t',x') = \frac{c_M}{3} (\partial_{t}\partial^2_{x} + \partial_{x}\partial_{t'}\partial_{x} - \partial_{t}\partial^2_{x}) \partial^2_{x'}\Big(\frac{1}{x - x'}\Big)
\end{eqnarray}
Using the delta function \eqref{deltafn-plane} and $\partial_{x}G = (x - x')^{-1}$, $\partial_{x'}G = -(x - x')^{-1}$, this gives
\begin{equation}
	T^{\alpha}_{\ \alpha}(t,x)T^{\beta}_{\ \beta}(t',x') = -\frac{2\pi c_M}{3} \partial_{x}\partial_{x'}\delta^{(2)}(\Delta t, \Delta x) , \label{traceope-plane}
\end{equation}
where $\Delta t = t - t'$ and $\Delta x = x - x'$. Under an infinitesimal Weyl transformation of the null plane $\delta e^a_{\alpha} =\Omega \delta^a_{\alpha}$, we have from \eqref{deltatrace}
\begin{equation}
	\delta \langle T^{\alpha}_{\ \alpha}(t,x)\rangle_{\delta} = \int dt'dx' 2\, T^{\alpha}_{\ \alpha}(t,x) T^{\beta}_{\ \beta}(t',x')\Omega(t',x').
\end{equation}
Substituting the $T^{\alpha}_{\ \alpha}(t,x)T^{\beta}_{\ \beta}(t',x')$ OPE \eqref{traceope-plane} and simplifying, we get
\begin{equation}
	\delta \langle T^{\alpha}_{\ \alpha}(t,x)\rangle_{\delta} = \frac{4\pi}{3}  c_M\partial^2_{x}\Omega.
\end{equation}
Further using \eqref{traceexpansion}, we get the trace anomaly in a 2d BMSFT on a slightly curved Carroll background described by $e^a_{\alpha}(t,x) = e^{\Omega(t,x)}\delta^a_{\alpha}$, with $\Omega(t,x)$ small, as
\begin{equation}
	\langle T^{\alpha}_{\ \alpha}\rangle_e = \frac{4\pi}{3} c_M\partial^2_{x}\Omega + (\text{higher order in}\ \Omega). \label{anomaly-plane}
\end{equation}

\bigskip

\subsection{Anomaly for null cylinder}
On the cylinder, we use the delta function identity \refb{deltafn-cylinder} and the form of the TT OPE \refb{TTope-cylinder}. Now to solve the differential equation \eqref{traceeqn}, we substitute the central terms in the OPEs \eqref{TTope-cylinder} in \eqref{traceeqn} to get
\begin{eqnarray}\label{traceeqn1}
	\partial_{\sigma} \partial_{\sigma'} T^{\alpha}_{\ \alpha}T'^{\beta}_{\ \beta} = 2c_M\left[ \partial_{\tau} \partial_{\tau'} \Big(\frac{-2\Delta\tau c_{\sigma\sigma'}}{s_{\sigma\sigma'}^4}+\frac{\Delta\tau c_{\sigma\sigma'} }{6s_{\sigma\sigma'}^2}\Big) -\partial_{\sigma} \partial_{\tau'} \Big (\frac{1}{s_{\sigma\sigma'}^4}-\frac{1}{6s_{\sigma\sigma'}^2}\Big) -\partial_{\sigma'} \partial_{\tau} \Big(\frac{1}{s_{\sigma\sigma'}^4}-\frac{1}{6s_{\sigma\sigma'}^2}\Big) \right]. \nonumber
\end{eqnarray}
where $s_{\sigma\sigma'} \equiv 2\sin(\frac{\sigma - \sigma'}{2})$, $c_{\sigma\sigma'}\equiv \cot(\frac{\sigma - \sigma'}{2})$, $\Delta\tau = \tau - \tau'$. Simplifying further using the Green's function in \eqref{deltafn-cylinder}, we get
\begin{eqnarray}
	\partial_{\sigma} \partial_{\sigma'} T^{\alpha}_{\ \alpha}(\tau,\sigma)T^{\beta}_{\ \beta}(\tau',\sigma') = \frac{c_M}{3} \left(-\partial_{\tau}\partial^{4}_{\sigma}\partial_{\sigma'}G + \partial_{\tau'}\partial^3_{\sigma}\partial^2_{\sigma'} G -\partial_{\tau}\partial^{3}_{\sigma}\partial^2_{\sigma'} G\right),
\end{eqnarray}
which upon using the delta function \eqref{deltafn-cylinder} gives the $T^{\alpha}_{\ \alpha}(\tau,\sigma)T^{\beta}_{\ \beta}(\tau',\sigma')$ OPE
\begin{eqnarray}
	T^{\alpha}_{\ \alpha}(\tau,\sigma)T^{\beta}_{\ \beta}(\tau',\sigma')=-\frac{2\pi}{3} c_M\partial_{\sigma}\partial_{\sigma'} \delta^{(2)}(\tau-\tau',\sigma-\sigma').
\end{eqnarray}
Substituting this OPE in the variation of the trace \eqref{deltatrace} under an infinitesimal BMS-Weyl transformation of the cylinder $\delta e^a_{\alpha} = \Omega \delta^a_{\alpha}$, we get 
\begin{eqnarray}
	\hspace{-6mm} \delta \langle T^{\alpha}_{\ \alpha}(\tau,\sigma)\rangle_{\delta} = -\frac{4\pi}{3}\int d^2\sigma' \big[ c_M\partial_{\sigma}\partial_{\sigma'} \delta^{(2)}(\tau-\tau',\sigma-\sigma') \big] \Omega(\tau',\sigma') = \frac{4\pi}{3} c_M\partial^2_{\sigma}\Omega(\tau,\sigma).
\end{eqnarray}
Using $\langle T^{\alpha}_{\ \alpha}(\tau,\sigma)\rangle_{\delta} = 0$, and the expression for $\delta \langle T^{\alpha}_{\ \alpha}(\tau,\sigma)\rangle_{\delta}$ above in \eqref{traceexpansion}, we get the trace anomaly in a 2d BMSFT on a slightly curved background described by $e^a_{\alpha}(\tau,\sigma) = e^{\Omega(\tau,\sigma)}\delta^a_{\alpha}$, to first order in $\Omega$:
\begin{equation}\label{BMS-Weyl-anomaly-cylinder}
	{\boxed{ \langle T^{\alpha}_{\ \alpha}(\tau,\sigma)\rangle_{e} = \frac{4\pi}{3} c_M\partial^2_{\sigma}\Omega(\tau,\sigma).}}
\end{equation}
This expression for the anomaly on the null cylinder \refb{BMS-Weyl-anomaly-cylinder} is the same as on the null plane. Our main result of this paper is the above expression for the BMS Weyl anomaly.

\subsection{Transformation of the partition function}
Now we turn to the transformation of partition functions under Weyl transformation. We first revisit the case for usual 2d CFTs, where the Liouville action arises as a response of Weyl variations to matter conformal theories. We then go on to considering 2d BMSFTs, where we will end up with a Carrollian version of Liouville theory, which is interestingly distinct from the BMS-Liouville theories that have been studied in the literature \cite{Barnich:2012rz, Barnich:2013yka}. 

\subsection*{Liouville theory arising from 2d CFT partition function}
Consider two metrics related to each other by 
\begin{equation}
\bar{g}_{\alpha \beta}=e^{2\Omega(x)}\delta_{\alpha \beta}.
\end{equation}
We choose to work with a flat Euclidean background and perturbations around it. Variation  of the metric with respect to the Weyl factor changes the partition function $Z[\bar{g},\Phi]$ as
\begin{eqnarray}
\frac{1}{Z}\frac{\partial Z}{\partial \Omega}  &=& \frac{1}{Z}\int \mathcal{D} \Phi e^{-S} \Big[-\frac{\partial S}{\partial \bar{g}_{\alpha \beta}}\frac{\partial \bar{g}_{\alpha \beta}}{\partial \Omega} \Big] = \frac{1}{Z} \int \mathcal{D} \Phi e^{-S} \big [-\sqrt{\bar{g}}T^{\alpha}_{\ \alpha}\big].
\end{eqnarray}
Now if we use the expression for trace anomaly given by equation \eqref{cft-trace}, then
\begin{eqnarray}
\frac{1}{Z}\frac{\partial Z}{\partial \Omega} &=& \sqrt{\bar{g}} \Big( \frac{c}{6 \pi} \bar{\mathcal{R}}-\mu \Big) = -\frac{c}{6 \pi}\sqrt{g} \Big(2\nabla^2 \Omega+\mu e^{2\Omega} \Big).
\end{eqnarray}
We can solve the above equation by integrating the partition function with respect to the Weyl factor. This relates the partition functions on different backgrounds which differ by the Weyl factor $\Omega(x)$. The solution is
\begin{equation}
Z[\bar{g}]=Z[g] \exp \Big[-\int d^2 \sigma \sqrt{g}\Big(\mu e^{2\Omega}-\frac{c}{6 \pi} g^{\alpha \beta} \partial_{\alpha} \Omega \partial_{\beta} \Omega \Big) \Big].
\end{equation}
The action inside the exponential  governs  the dynamics for the Weyl factor. With some redefinitions of the parameters we can write the above as
\be{}
Z[\bar{g}]=Z[g] \exp S_L,
\ee
where the Liouville action $S_L$ is given by
\begin{equation} \label{Lville}
\mathcal{S}_L(\Omega,g)=\int d^2\sigma \sqrt{g} \,\Big\{\frac{c}{6 \pi} g^{\alpha \beta}\partial_{\alpha} \Omega \partial_{\beta} \Omega-\Lambda e^ {2\Omega }\Big\}.
\end{equation}
Here the second term in the action is the Liouville potential.

\subsection*{Carrollian Liouville theory}
We now study the transformation of the BMSFT partition function under a Weyl transformation. Let $Z[e]$ be the partition function for a 2d BMSFT, with an action $S[\phi,e^a_{\alpha}]$, on a Carrollian background described by $e^a_{\alpha}(\tau,\sigma)$. Then under an infinitesimal BMS-Weyl transformation $e'^a_{\alpha} =e^{\Omega}e^a_{\alpha}$; $\delta e'^a_{\alpha} =\Omega e^a_{\alpha}$, the partition function changes as
\begin{eqnarray}
	\frac{1}{Z}\frac{\delta Z}{\delta \Omega} &=& \frac{1}{Z}\int \mathcal{D}\phi e^{-S[\phi,e'^a_{\alpha}]}\Big(-\frac{\delta S}{\delta\Omega}\Big) = \frac{1}{Z}\int \mathcal{D}\phi e^{-S[\phi,e'^a_{\alpha}]}\Big(-\frac{\delta S}{\delta e'^{\alpha}_a}\frac{\delta e'^{\alpha}_a}{\delta\Omega}\Big) \nonumber \\
	&=& \frac{1}{Z}\int \mathcal{D}\phi e^{-S[\phi,e'^a_{\alpha}]}\Big(2\,e'\,T'^{\alpha}_{\ \ \alpha}\big) =2\,e'\langle T^{\alpha}_{\ \alpha}\rangle_{e'}.
\end{eqnarray}
Choosing $e^a_{\alpha} = \delta^a_{\alpha}$, \ie\ a cylinder gives $e' = e^{\Omega}$, and using the expression for the anomaly \eqref{BMS-Weyl-anomaly-cylinder}, we get
\begin{equation}
	\frac{\delta}{\delta \Omega}\log Z = \frac{8\pi}{3} c_M\partial_{\sigma}^2\Omega.
\end{equation}
This change in $Z$ can be obtained by variation of a local action with respect to $\Omega(\tau,\sigma)$ as
\begin{equation}
	\log Z[e'] - \log Z[e] = \frac{8\pi}{3} c_M \int d^2\sigma \, \partial_{\sigma}\Omega\partial_{\sigma}\Omega,
\end{equation}
which can also be written as
\begin{equation}\label{BMSLiouville}
	Z[e'] = Z[e] e^{S_{cL}}\ ; \qquad S_{cL} = \frac{8\pi}{3} c_M \int d^2\sigma \, \partial_{\sigma}\Omega\partial_{\sigma}\Omega.
\end{equation}
The action $S_L$ is the Carrollian analogue of the Liouville action \eqref{Lville} which arises in relativistic CFTs. It can be shown that this action is invariant under BMS$_3$ transformations \cite{inprogress}. Like in the usual 2d CFT case, where the potential term appears due to renormalisation, one can envision that a similar term may arise in this context as well. In this case, the action would take the form 
\be{}
\hat{S}_{cL} = \int d^2\sigma \big( c_M\partial_{\sigma}\Omega\partial_{\sigma}\Omega - \mu e^{2\Omega} \big).
\ee
It is interesting to note that the action here is different from the ones obtained by the flat limit of the Liouville CFT. Liouville theory has been used in the boundary dynamics of AdS$_3$ at the classical level \cite{Barnich:2012rz, Barnich:2013yka}. The flat versions have been used to provide a similar holographic description of 3d flat spacetime. It would be of interest to figure out what this new version leads to in terms of flat holography.


\section{Conclusions}
\subsection*{Summary}
In this paper, we have provided a first derivation for a quantum anomaly in a BMS invariant field theory. Specifically, we have found the form of the Weyl anomaly or equivalently the trace anomaly in 2d BMSFTs. For this we needed the form of the OPEs of the stress tensors defined in these field theories and importantly a delta-function identity for the Carrollian manifolds on which these theories are defined. We proposed such an identity for the BMSFTs defined on a null plane and showed that the identity correctly reproduced the stress energy correlation functions. While working on the null cylinder, we used an identity proposed earlier in \cite{Bagchi:2015wna}, which had also produced $n$-point correlation functions of the BMS EM tensors, which were also checked by a limit from relativistic answers as well as matched with a holographic computation in 3d flat space using the Chern-Simons formulation. Our calculations on the null cylinder yielded more general results, for e.g. the Ward identities, as unlike in \cite{Bagchi:2015wna}, we chose to work with non-vanishing $c_L$. For purposes of computing the Weyl anomaly, without contamination from potential diffeomorphism anomalies, we then worked with $c_L=0$. The final expression for the BMS Weyl anomaly for both the plane and the cylinder for small perturbations away from a flat Carrollian manifold gives: 
\be{}
\langle T^{\alpha}_{\ \alpha}(\tau,\sigma)\rangle_{e} = \frac{4\pi}{3} c_M\partial^2_{\sigma}\Omega(\tau,\sigma).
\ee
Finally, we looked at the change in the BMS partition functions under Weyl transformations and found a Carrollian version of the Liouville action \refb{BMSLiouville}. This was different from the ones found earlier in e.g. \cite{Barnich:2012rz}. 

Our analysis would be of use not only for studies of Minkowskian holography, through these 2d BMSFTs, but to non-relativistic systems due to the duality between Galilean and Carrollian systems in $d=2$, and also to the study of tensionless strings. 

\subsection*{Discussions}
There are several directions of research that require immediate attention. First and foremost, it would be important to understand geometrically what the form of the BMS-Weyl anomaly is and whether it can be given a form similar to (the second equality of) \refb{cft-anom}. Our present investigations indicate that one can indeed give an answer analogous to \refb{cft-anom} in terms of the geometric data of the Carrollian manifold. This will be presented in upcoming work. 

\medskip
\noindent
Our principle follow-up work would be to generalise our construction to theories with $c_L\neq0$. As described in the paper, these theories would be parity violating and would have a diffeomorphism anomaly, which in turn would lead to a non-conservation of our EM tensor. In relativistic theories, this anomaly can be traded for a Lorentz anomaly making the EM tensor asymmetric. In our non-Lorentzian boundary theory, the EM tensor is already asymmetric, so perhaps the natural avenue to take is to ensure the conservation of the EM tensor and figure out where this leads and whether the resulting ``improved" stress tensor has something akin to Lorentz anomaly, perhaps a Carroll anomaly. We wish to also investigate consistent (as in consistent with Wess-Zumino conditions) and covariant versions of the Weyl anomaly to figure out potential changes in our formulae. 

\medskip
\noindent It would also be good to re-derive our results in other methods, e.g. using the Fujikawa procedure of path integrals. These methods would all need to be re-examined however as we have an underlying degenerate metric structure and a non-Riemmannian manifold where these field theories live. We are also interested in a holographic check of our anomaly using methods similar to \cite{Henningson:1998gx}, now for 3d asymptotically flat spacetimes. Some of the above are works in progress.

\medskip
\noindent Our version of the Carrollian Liouville action throws up very interesting avenues of further research. The difference with the limiting analysis of \cite{Barnich:2012rz} is intriguing. It would mean that there is potentially another limit that yields the Carrollian Liouville action. (We try and distinguish it from the earlier work in the literature by calling this Carrollian Liouville as opposed to BMS Liouville.) It would be instructive to construct this limit explicitly in the relativistic Liouville theory and understand what it means. It would also be very interesting to figure out how one can modify the analysis of \cite{Barnich:2013yka} to get the Carrollian Liouville theory starting out from a Chern-Simons formulation of 3d asymptotically flat gravity, then going over to a WZW model and then doing a Hamiltonian reduction. 

\medskip
\noindent 
Finally, it would be of interest to see how our results would fit in with non-relativistic anomalies. The algebra and the underlying field theory stays the same. Hence the answers, obtained from an intrinsic analysis, like we have in this paper, are supposed to hold, upto identifications of the space and time directions. But clearly there is something very interesting and mysterious happening here. The identification of central charges are flipped in the NR limit: 
\be{}
c_L=c+\bar{c}, \quad c_M = \e(c-\bar{c}). 
\ee
It seems that requiring no diffeomorphism anomaly would lead to $c_M=0$, but as stated earlier, this reduces the symmetries to a single copy of Virasoro algebra and the theory to a chiral half of a 2d CFT. This is clearly chiral and hence should be associated with a gravitational anomaly. It is hence far from clear what a gravitational anomaly means in the NR limit, and consequently what the correct form of the Weyl anomaly should be. We wish to investigate this in the near future with the hope of clarifying the above and other related confusions.

\section*{Acknowledgements}
It is a pleasure to thank Shamik Banerjee, Rudranil Basu, Daniel Grumiller, Jelle Hartong, and Marika Taylor for several illuminating discussions and comments on a draft of the paper. 

\smallskip 

\noindent AB's research is supported by a Swarnajayanti fellowship of the Department of Science and Technology and the Science and Engineering Research Board (SERB), India. AB is further supported by the following grants from SERB: EMR/2016/008037, ERC/2017/000873, MTR/2017/000740; as well as an international exchange grant (with the University of Southampton) from the Royal Society of London. 
\newpage


\appendix
\section*{APPENDICES} 
\section{Ward identities and correlators in BMSFTs: Details}

\subsection{BMSFT on null plane}\label{sec:Wid-plane-detials}
Here we provide some details of the computation of the Ward identity \eqref{Wid-plane-N} on the null plane. The time derivative of $\langle T_1\rangle_{\mu}$ in the $\mu$-deformed theory is
\begin{equation}
	\partial_t \langle T_1\rangle_{\mu} = \Big\langle \int dt'dx' \Big(\mu'_L \partial_t (T_1T'_1) + \mu'_M \partial_t (T_1T'_2) \Big) + \partial_t T_1 \Big\rangle_{\mu}.
\end{equation}
Using the OPEs \eqref{TTbms}, this simplifies to
\begin{eqnarray}
	\partial_t \langle T_1\rangle_{\mu} &=& \Big\langle \int dt'dx' \Big(\mu'_L \partial_t \Big[\frac{c_L}{2(\Delta x)^4}-\frac{2c_M(\Delta t)}{(\Delta x)^5} + \frac{2T'_1}{(\Delta x)^2} -\frac{4T'_2(\Delta t)}{(\Delta x)^3} +\frac{\partial_{x'}T'_1}{(\Delta x)}-\frac{\partial_{t'}T'_1(\Delta t)}{(\Delta x)^2} \Big] \nonumber \\
	&& \qquad  + \mu'_M \partial_t\Big[\frac{c_M}{2(\Delta x)^4} + \frac{2T'_2}{(\Delta x)^2} +\frac{\partial_{x'}T'_2}{(\Delta x)}\Big] \Big) + \partial_t T_1 \Big\rangle_{\mu}\ , \nonumber \\
	&=& \Big\langle \int dt'dx' \Big(\mu'_L\Big[\frac{c_L}{12}\partial^2_x\partial_x'\partial_t\Big(\frac{1}{\Delta x}\Big) -2T'_1\partial_x\partial_t\Big(\frac{1}{\Delta x}\Big) +\partial_x'T'_1\partial_t\Big(\frac{1}{\Delta x}\Big) \Big] \nonumber \\
	&& \qquad + \mu'_M\Big[\frac{c_M}{12}\partial^2_x\partial_x'\partial_t\Big(\frac{1}{\Delta x}\Big) - 2T'_2\partial_x\partial_t\Big(\frac{1}{\Delta x}\Big)+\partial_x'T'_2\partial_t\Big(\frac{1}{\Delta x}\Big) \Big] \nonumber \\
	&& \qquad + \mu'_L\Big[-\frac{2c_M}{\Delta x^5}-\frac{4T'_2}{\Delta x^3}-\frac{\partial_t'T'_1}{\Delta x^2} \Big] + \partial_t T_1 \nonumber \\
	&& \qquad + \Delta t \mu'_L\Big[-2c_M\partial_t\Big(\frac{1}{\Delta x^5}\Big) -4T'_2\partial_t\Big(\frac{1}{\Delta x^3}\Big) -\partial_t'T'_1\partial_t\Big(\frac{1}{\Delta x}\Big) \Big] \Big) \Big\rangle_{\mu}. \label{deltT1-plane}
\end{eqnarray}
Using the delta function \eqref{deltafn-plane}, we see that each term in the $4th$ line above is proportional to $\int dt'dx' \Delta t \delta^{(2)}(\Delta t,\Delta x) \sim 0$. We can write the $3rd$ line above in terms of $\mathcal{M}$ as
\begin{eqnarray}
	\partial_x \mathcal{M} = \partial_x\langle T_2\rangle_{\mu} &=& \Big\langle \int dt'dx'\mu'_L\partial_x(T_2T'1) + \partial_x T_2 \Big\rangle_{\mu} \nonumber \\
	&=& \Big\langle \int dt'dx'\mu'_L \Big[-\frac{2c_M}{\Delta x^5}-\frac{4T'_2}{\Delta x^3}-\frac{\partial_t'T'_1}{\Delta x^2} \Big] + \partial_x T_2 \Big\rangle_{\mu}.
\end{eqnarray}
Substituting this expression for $\partial_x \mathcal{M}$ and using the conservation equation $\partial_tT_1 = \partial_xT_2$, \eqref{deltT1-plane} simplifies to
\begin{equation}
	\frac{1}{2\pi}(\partial_t\langle T_1\rangle_{\mu} - \partial_x \mathcal{M}) = \Big\langle  -\frac{c_L}{12}\partial^3_x\mu_L -2\partial_x(\mu_L T_1) +\mu_L\partial_xT_1 -\frac{c_M}{12}\partial^3_x\mu_M-2\partial_x(\mu_MT_2) +\mu_M\partial_xT_2 \Big\rangle_{\mu},
\end{equation}
which upon using $\mathcal{N} = \langle T_1\rangle_{\mu}$ and $\mathcal{M} = \langle T_2\rangle_{\mu}$ is written as \eqref{Wid-plane-N1}.

\subsection{BMSFT on null cylinder}\label{sec:Wid-cylinder}

The derivation of correlators of the stress tensor components from Ward identities using the delta function \eqref{deltafn-cylinder} was given in \cite{Bagchi:2015wna} for 2d BMSFTs on cylinder with $c_L = 0$. Here we generalize the analysis for $c_L \neq 0$. Since the details of this analysis are same as in \ref{sec:Wid-plane}, we present only key steps and results here.

We consider a deformation to a free 2d BMSFT on a null cylinder, described by an action $S_0$, by localized sources for the stress tensor components\,:
\begin{eqnarray}
	&& S_{\mu} = S_0 -\int d\tau d\sigma (\mu_L(\tau,\sigma) T_1(\tau,\sigma) + \mu_M(\tau,\sigma) T_2(\tau,\sigma)); \label{deformed-action-cylinder} \\
	&& \mu_L(\tau,\sigma) = \epsilon_L\delta^{(2)}(\tau-\tau',\sigma-\sigma')\ , \qquad \mu_M(\tau,\sigma) = \epsilon_M\delta^{(2)}(\tau-\tau',\sigma-\sigma'). \label{mu-sources-cylinder}
\end{eqnarray}
The expectation values of $T_1$, $T_2$ in the deformed theory are related to the correlators of $T_1$, $T_2$ in the free theory as
\begin{eqnarray}
	&& \hspace{-9mm} \langle T_1(\tau,\sigma)\rangle_{\mu} = \langle T_1(\tau,\sigma)\rangle_0 + \epsilon_L\langle T_1(\tau,\sigma) T_1(\tau',\sigma')\rangle_0 + \epsilon_M\langle T_1(\tau,\sigma) T_2(\tau',\sigma')\rangle_0 + O(\epsilon^2), \\
	&& \hspace{-9mm} \langle T_2(\tau,\sigma)\rangle_{\mu} = \langle T_2(\tau,\sigma)\rangle_0 + \epsilon_L\langle T_2(\tau,\sigma) T_1(\tau',\sigma')\rangle_0 + \epsilon_M\langle T_2(\tau,\sigma) T_2(\tau',\sigma')\rangle_0 + O(\epsilon^2) .
\end{eqnarray}
Defining $\mathcal{M} = \langle T_2\rangle_{\mu}$, $\mathcal{N} = \langle T_1\rangle_{\mu}$ and expanding in $\epsilon_{L/M}$ as
\begin{equation}
	\mathcal{M} = \mathcal{M}^{(0)} + \mathcal{M}^{(1)} + \cdots, \qquad \mathcal{N} = \mathcal{N}^{(0)} + \mathcal{N}^{(1)} + \cdots, \label{MN-expansion-cylinder}
\end{equation}
where $\mathcal{M}^{(n)} \sim O(\epsilon^n)$ and so on, we get
\begin{eqnarray}
	&& \mathcal{M}^{(0)} = \langle T_2\rangle_0, \qquad \mathcal{M}^{(1)} = \epsilon_L\langle T_2T'_1\rangle_0 +\epsilon_M\langle T_2T'_2\rangle_0, \nonumber \\
	&& \mathcal{N}^{(0)} = \langle T_1\rangle_0, \qquad \mathcal{N}^{(1)} = \epsilon_L\langle T_1T'_1\rangle_0 +\epsilon_M\langle T_1T'_2\rangle_0. \label{M1N1-correlators}
\end{eqnarray}
Here we have reverted to the shorthand $T_1 = T_1(\tau,\sigma)$, $T'_1 = T_1(\tau',\sigma')$,
Now we take the time derivative of $\langle T_2\rangle_{\mu}$, and use $\partial_{\tau}S_0 = 0$, a conservation equation $\partial_{\tau}T_2 = 0$ and the OPEs \eqref{TTope-cylinder} to get
\begin{eqnarray}
	\partial_{\tau} \langle T_2\rangle_{\mu} &=& \Big\langle \int d\tau'd\sigma' \Big(\mu'_L \partial_{\tau} (T_2T'_1) + \mu'_M \partial_{\tau} (T'_2 T_2) \Big)\Big\rangle_{\mu} \\
	&=& \Big\langle \int d\tau'd\sigma'\mu'_L \partial_{\tau}\Big[\frac{c_M}{2\big(2\sin\frac{\Delta\sigma}{2}\big)^4} - \frac{c_M}{12\big(2\sin\frac{\Delta\sigma}{2}\big)^2} + \frac{2T'_2}{\big(2\sin\frac{\Delta\sigma}{2}\big)^2} + \frac{\partial_{\sigma'}T'_2}{2\sin\frac{\Delta\sigma}{2}}\Big] \Big\rangle_{\mu}, \nonumber
\end{eqnarray}
where $\Delta\sigma = \sigma - \sigma'$ and $\Delta\tau = \tau-\tau'$. Using the delta function \eqref{deltafn-cylinder}, we can write
\begin{equation}
	\partial_{\sigma'}\partial^4_{\sigma}G + \partial_{\sigma'}\partial^2_{\sigma}G = -\frac{3}{4}\frac{\cos\frac{\Delta\sigma}{2}}{\big(\sin\frac{\Delta\sigma}{2}\big)^5}, \quad \partial_{\tau}\Big(\frac{1}{2\sin\frac{\Delta\sigma}{2}}\Big) = \partial_{\tau}\Big(\partial_{\sigma}G + \frac{1}{2}\tan\frac{\Delta\sigma}{4}\Big) = \partial_{\tau}\partial_{\sigma}G,
\end{equation}
since $\tan\frac{\Delta\sigma}{4}$ is non-singular as $\Delta\sigma\rightarrow 0$. Using these expressions and the delta function \eqref{deltafn-cylinder}, $\partial_{\tau} \langle T_2\rangle_{\mu}$ simplifies to
\begin{eqnarray}
	\partial_{\tau}\langle T_2\rangle_{\mu} &=& \Big\langle \int d\tau'd\sigma'\mu'_L\partial_{\tau}\Big[\frac{c_M}{12}\partial^2_{\sigma}\partial_{\sigma'}(\partial_{\sigma}G) + 2T'_2\partial_{\sigma'}(\partial_{\sigma}G) + \partial_{\sigma'}T'_2(\partial_{\sigma}G)\Big]\Big\rangle_{\mu} \nonumber \\
	&=& \Big\langle 2\pi \int d\tau'd\sigma' \mu'_L \Big[\frac{c_M}{12}\partial^2_{\sigma}\partial_{\sigma'}\delta^{(2)} + 2T'_2\partial_{\sigma'}\delta^{(2)}  + \delta^{(2)}\partial_{\sigma'}T'_2\Big] \Big\rangle_{\mu} \nonumber \\
	&=& \Big\langle 2\pi \Big( -\frac{c_M}{12}\partial^3_{\sigma}\mu_L -\partial_{\sigma}(2T_2\mu_L) + \partial_{\sigma}T_2\mu_L \Big) \Big\rangle_{\mu}, \nonumber \\
	\Rightarrow -\frac{1}{2\pi}\partial_{\tau}\mathcal{M} &=& \frac{c_M}{12}\partial^3_{\sigma}\mu_L + 2\mathcal{M}\partial_{\sigma}\mu_L + \mu_L\partial_{\sigma}\mathcal{M}. \label{Wid-cylinder-M}
\end{eqnarray}
Similarly
\begin{equation}
	\partial_{\tau} \langle T_1\rangle_{\mu} = \Big\langle \int d\tau'd\sigma'\big(\mu'_L\partial_{\tau}(T_1T'_1) + \mu'_M\partial_{\tau}(T_1T'_2) \big) + \partial_{\tau}T_1 \Big\rangle_{\mu}. \label{deltauT1-cylinder}
\end{equation}
The $\mu_M$ term in \eqref{deltauT1-cylinder} is same as the $\mu_L$ term in $\partial_{\tau}\langle T_2\rangle_{\mu}$ above. Let us simplify the $\mu_L$ term\,:
\begin{eqnarray}
	&& \int d\tau'd\sigma'\mu'_L \partial_{\tau}\Big[\frac{c_L}{2(2\sin(\frac{\Delta\sigma}{2}))^4} -\frac{c_L}{12(2\sin(\frac{\Delta\sigma}{2}))^2}-\frac{2c_M(\Delta\tau)\cos(\frac{\Delta\sigma}{2})}{(2\sin(\frac{\Delta\sigma}{2}))^5} +\frac{2c_M(\Delta\tau)\cos(\frac{\Delta\sigma}{2})}{12(2\sin(\frac{\Delta\sigma}{2}))^3} \nonumber \\
	&& \qquad +\frac{2T_1(\tau',\sigma')}{(2\sin(\frac{\Delta\sigma}{2}))^2}-\frac{4T_2(\tau',\sigma')\cos(\frac{\Delta\sigma}{2})}{(2\sin(\frac{\Delta\sigma}{2}))^3} +\frac{\partial_{\sigma'}T_1(\tau',\sigma')}{2\sin(\frac{\Delta\sigma}{2})} -\frac{\partial_{\sigma'}T_2(\tau',\sigma')(\Delta\tau)}{(2\sin(\frac{\Delta\sigma}{2}))^2} \Big] \nonumber \\
	= && \int d\tau'd\sigma' \Big(\mu'_L \partial_{\tau}\Big[\frac{c_L}{12}\partial^2_{\sigma}\partial_{\sigma'}(\partial_{\sigma}G) +2T'_1\partial_{\sigma'}(\partial_{\sigma}G) + \partial_{\sigma'}T'_1(\partial_{\sigma}G)\Big] \nonumber\\
	&& \qquad +\mu'_L\Big[2c_M\Big(-\frac{\cos\frac{\Delta\sigma}{2}}{\big(2\sin\frac{\Delta\sigma}{2}\big)^5} + \frac{\cos\frac{\Delta\sigma}{2}}{12\big(2\sin\frac{\Delta\sigma}{2}\big)^3} \Big) -\frac{4T'_2\cos\frac{\Delta\sigma}{2}}{\big(2\sin\frac{\Delta\sigma}{2}\big)^3} - \frac{\partial_{\sigma'}T'_2\cos\frac{\Delta\sigma}{2}}{\big(2\sin\frac{\Delta\sigma}{2}\big)^2} \Big] \\
	&& \qquad +\Delta\tau \mu'_L\partial_{\tau}\Big[2c_M\Big(-\frac{\cos\frac{\Delta\sigma}{2}}{\big(2\sin\frac{\Delta\sigma}{2}\big)^5} + \frac{\cos\frac{\Delta\sigma}{2}}{12\big(2\sin\frac{\Delta\sigma}{2}\big)^3} \Big) -\frac{4T'_2\cos\frac{\Delta\sigma}{2}}{\big(2\sin\frac{\Delta\sigma}{2}\big)^3} - \frac{\partial_{\sigma'}T'_2\cos\frac{\Delta\sigma}{2}}{\big(2\sin\frac{\Delta\sigma}{2}\big)^2} \Big] \Big). \nonumber
\end{eqnarray}
The $3rd$ line in the above expression is proportional to $\int d\tau'd\sigma' \Delta\tau \delta^{(2)}(\Delta\tau,\Delta\sigma) \sim 0$ and the $2nd$ line can be expressed in terms of $\mathcal{M}$ as
\begin{eqnarray}
	\partial_{\sigma}\mathcal{M} &=& \Big\langle \int d\tau'd\sigma'\big(\mu'_L\partial_{\sigma}(T_2T'_1) + \mu'_M\partial_{\sigma}(T_2T'_2) \big) + \partial_{\sigma}T_2 \Big\rangle_{\mu} \nonumber \\
	&=& \Big\langle \int d\tau'd\sigma' \mu'_L \partial_{\sigma}\Big[ \frac{c_M}{2(2\sin(\frac{\Delta\sigma}{2}))^4}-\frac{c_M}{12(2\sin(\frac{\Delta\sigma}{2}))^2}+\frac{2T_2(\tau',\sigma')}{(2\sin(\frac{\Delta\sigma}{2}))^2} \nonumber \\
	&& \qquad  +\frac{(\partial_{\tau'}T_1(\tau',\sigma')+\partial_{\sigma'}T_2(\tau',\sigma'))}{4\sin(\frac{\Delta\sigma}{2})} \Big] + \partial_{\sigma}T_2 \Big\rangle_{\mu} \nonumber \\
	&=& \Big\langle \int d\tau'd\sigma' \mu'_L\Big[2c_M\Big(-\frac{\cos\frac{\Delta\sigma}{2}}{\big(2\sin\frac{\Delta\sigma}{2}\big)^5} + \frac{\cos\frac{\Delta\sigma}{2}}{12\big(2\sin\frac{\Delta\sigma}{2}\big)^3} \Big) -\frac{4T'_2\cos\frac{\Delta\sigma}{2}}{\big(2\sin\frac{\Delta\sigma}{2}\big)^3} \nonumber \\
	&& \qquad  - \frac{\partial_{\sigma'}T'_2\cos\frac{\Delta\sigma}{2}}{\big(2\sin\frac{\Delta\sigma}{2}\big)^2} \Big] + \partial_{\sigma}T_2 \Big\rangle_{\mu}\ .
\end{eqnarray}
Then using these expressions, we get
\begin{eqnarray}
	\partial_{\tau}\langle T_1 \rangle_{\mu} &=& \Big\langle 2\pi \Big( - \frac{c_L}{12}\partial^3_{\sigma}\mu_L - \partial_{\sigma}(2T_1\mu_L) + \mu_L\partial_{\sigma}T_1  \nonumber \\
	&& \qquad - \frac{c_M}{12}\partial^3_{\sigma}\mu_M - 2T_2\partial_{\sigma}\mu_M - \mu_M\partial_{\sigma}T_2 \Big) \Big\rangle_{\mu} +\partial_{\sigma}\mathcal{M}, \nonumber \\
	\ie\quad -\frac{1}{2\pi}(\partial_{\tau}\mathcal{N}-\partial_{\sigma}\mathcal{M}) &=& \frac{c_L}{12}\partial^3_{\sigma}\mu_L + 2\mathcal{N}\partial_{\sigma}\mu_L + \mu_L\partial_{\sigma}\mathcal{N} \nonumber \\
	&& \ +\frac{c_M}{12}\partial^3_{\sigma}\mu_M + 2\mathcal{M}\partial_{\sigma}\mu_M + \mu_M\partial_{\sigma}\mathcal{M}. \label{Wid-cylinder-N}
\end{eqnarray}
Now expanding the Ward identities \eqref{Wid-cylinder-M},\eqref{Wid-cylinder-N} using the expansion \eqref{MN-expansion-cylinder} and the expression for the sources \eqref{mu-sources-cylinder}, we get\,:\\
$(i)$ conservation equations from the leading term
\begin{equation}
	\partial_{\tau}\mathcal{M}^{(0)} = 0\ , \qquad \partial_{\tau}\mathcal{N}^{(0)} = \partial_{\sigma}\mathcal{M}^{(0)},
\end{equation}
and \newline $(ii)$ Ward identities for $\mathcal{M}^{(1)}$ and $\mathcal{N}^{(1)}$ from $O(\epsilon)$ terms
\begin{eqnarray}
	-\frac{1}{2\pi}\partial_{\tau} \mathcal{M}^{(1)} &=& \epsilon_L\Big[\frac{c_M}{12}\partial^3_{\sigma}\delta^{(2)} +2\mathcal{M}^{(0)}\partial_{\sigma}\delta^{(2)} +\delta^{(2)}\partial_{\sigma}\mathcal{M}^{(0)} \Big], \label{Wid-cylinder-M1} \\
	- \frac{1}{2\pi}(\partial_{\tau}\mathcal{N}^{(1)} - \partial_{\sigma}\mathcal{M}^{(1)}) &=& \epsilon_L\Big[\frac{c_L}{12}\partial^3_{\sigma}\delta^{(2)} +2\mathcal{N}^{(0)}\partial_{\sigma}\delta^{(2)} +\delta^{(2)}\partial_{\sigma}\mathcal{N}^{(0)} \Big] \nonumber \\
	&& + \epsilon_M\Big[\frac{c_M}{12}\partial^3_{\sigma}\delta^{(2)} +2\mathcal{M}^{(0)}\partial_{\sigma}\delta^{(2)} +\delta^{(2)}\partial_{\sigma}\mathcal{M}^{(0)} \Big]. \label{Wid-cylinder-N1}
\end{eqnarray}
Using the values $\mathcal{M}^{(0)} = \frac{c_M}{24}$ and $\mathcal{N}^{(0)} = \frac{c_L}{24}$ in a free BMS$_3$ invariant field theory on a cylinder and the delta function \eqref{deltafn-cylinder}, we solve the above Ward identities to get
\begin{equation}
	\mathcal{M}^{(1)} = \epsilon_L\frac{c_M}{2\big(2\sin\frac{\Delta\sigma}{2}\big)^4}, \quad \mathcal{N}^{(1)} = \epsilon_L\Big(\frac{c_L}{2\big(2\sin\frac{\Delta\sigma}{2}\big)^4} - \frac{2c_M\Delta\tau\cos\frac{\Delta\sigma}{2}}{\big(2\sin\frac{\Delta\sigma}{2}\big)^5} \Big) + \epsilon_M\frac{c_M}{2\big(2\sin\frac{\Delta\sigma}{2}\big)^4} .
\end{equation}
Comparing these with \eqref{M1N1-correlators}, we get the desired correlators as
\begin{equation}
	\langle T_1T'_1\rangle = \frac{c_L}{2\big(2\sin\frac{\Delta\sigma}{2}\big)^4} - \frac{2c_M\Delta\tau\cos\frac{\Delta\sigma}{2}}{\big(2\sin\frac{\Delta\sigma}{2}\big)^5}, \quad \langle T_1T'_2\rangle = \frac{c_M}{2\big(2\sin\frac{\Delta\sigma}{2}\big)^4}, \quad \langle T_2T'_2\rangle = 0 .
\end{equation}
The above method can be generalised to obtain arbitrary $N$-point functions of the BMS EM tensors. For $c_L=0$, the results were also verified holographically by a computation in 3d asymptotically flat Einstein gravity. In order to reproduce the results with a non-zero $c_L$, one would need to look at a generalisation of the bulk calculation to theories of topologically massive gravity where one can generate two non-zero central charges. 

\bigskip \bigskip

\section{Zweibein formulation}\label{sec:zweibeinformulation}
The zweibein formulation of Carrollian geometries is discussed in \cite{Hartong:2015xda,Hartong:2015usd,Bergshoeff:2017btm}. In the zweibein formulation, a $(1+1)$ dimensional Carrollian geometry is described by the zweibeins $e^0_{\alpha}(\tau,\sigma)$ and $e^1_{\alpha}(\tau,\sigma)$, with their inverses $e^{\alpha}_0(\tau,\sigma)$ and $e^{\alpha}_1(\tau,\sigma)$ defined through the relations
\begin{equation}
	e^0_{\alpha}e^{\alpha}_0 = 1, \quad e^1_{\alpha}e^{\alpha}_1 = 1, \quad e^0_{\alpha}e^{\alpha}_1 = 0, \quad e^1_{\alpha}e^{\alpha}_0 = 0, \quad e^0_{\alpha}e^{\beta}_0 + e^1_{\alpha}e^{\beta}_1 = \delta^{\beta}_{\alpha}.
\end{equation}
For our discussion, we collectively write $e^0_{\alpha}$ and $e^1_{\alpha}$ as $e^a_{\alpha}$ and the inverses as $e^{\alpha}_a$, in terms of which the above relations become
\begin{equation}
	e^a_{\alpha}e^{\alpha}_b = \delta^a_b, \qquad e^a_{\alpha}e^{\beta}_a = \delta^{\beta}_{\alpha}.
\end{equation}
Here $\alpha,\beta$ are spacetime indices denoting the coordinates $(\tau,\sigma)$ and $a,b,\dots = 0,1$ are tangent space indices. We introduce antisymmetric symbols $\epsilon^{\alpha\beta}$ and $\epsilon^{ab}$ defined as $\epsilon^{\tau\sigma}=1=-\epsilon_{\tau\sigma}$ and $\epsilon^{01}=1=-\epsilon_{01}$, which satisfy $\epsilon_{\alpha\rho}\epsilon^{\rho\beta}=\delta_{\alpha}^{\beta}$ and $\epsilon_{ac}\epsilon^{cb}=\delta_a^b$. Using these antisymmetric symbols, we can write the determinants $e=\det(e^a_{\alpha})$, $\frac{1}{e}=\det(e^{\alpha}_a)$ and inverse zweibeins as
\begin{equation}
	e=\epsilon^{\alpha\beta}e^0_{\alpha}e^1_{\beta}=\frac{1}{2}\epsilon_{ab}e^a_{\alpha}e^b_{\beta}\epsilon^{\beta\alpha}, \qquad \frac{1}{e} = \frac{1}{2}\epsilon^{ab}e^{\alpha}_a e^{\beta}_b\epsilon_{\beta\alpha}, \qquad e^{\alpha}_a = \frac{1}{e}\epsilon^{\alpha\beta}e^b_{\beta}\epsilon_{ba}.
\end{equation}
Under spacetime diffeomorphisms $\xi^{\alpha}\rightarrow \xi'^{\alpha}(\xi)$, the zweibeins transform as
\begin{equation}
	e'^{\alpha}_a = \frac{\partial\xi'^{\alpha}}{\partial\xi^{\beta}}e^{\beta}_a.
\end{equation}

\medskip

\noindent A flat Carrollian spacetime in $(1+1)$ dimensions, with a degenerate metric, given by
\begin{equation}\label{flatCarroll-metric}
	ds^2=g_{\alpha\beta}d\sigma^{\alpha}d\sigma^{\beta} = -0\cdot d\tau^2 + d\sigma^2, \qquad \zeta = \frac{\partial}{\partial\tau}, \qquad \Gamma^{\alpha}_{\beta\rho} = 0,
\end{equation}
is described by zweibeins
\begin{equation}
	e^a_{\alpha} = \delta^a_{\alpha}, \qquad e^{\alpha}_a = \delta^{\alpha}_a. \label{flatCarroll-zweibeins}
\end{equation}

\bigskip

\subsection*{Carroll-Weyl and BMS-Weyl transformations}

In \cite{Ciambelli:2018wre,Ciambelli:2018xat,Ciambelli:2019lap} (see also \cite{Gupta:2020dtl}), Carrollian geometry is described in the metric formulation. In this formulation in \cite{Ciambelli:2019lap}, Carroll-Weyl transformations are defined such that the metric data describing the temporal part of the geometry scales differently than the non-degenerate metric on the spatial hypersurface, the difference in scaling being governed by a real number $z$. For $z = \frac{2}{N}$, for integer $N$ the conformal isometries of the Carrollian spacetime form the conformal Carroll algebra $\mathfrak{ccarr}_{N}(d+1)$ of level $N$ \cite{Duval:2014uoa,Duval:2014uva,Duval:2014lpa}. It was shown that for $N=2$ ($z=1$), the conformal Carroll algebra $\mathfrak{ccarr}_{2}(d+1)$ is isomorphic to BMS algebra in $d+2$ dimensions \ie\ $\mathfrak{ccarr}_{2}(d+1) \simeq \mathfrak{bms}_{(d+2)}$. In particular, for our case of $d=1$, we have $\mathfrak{ccarr}_{2}(2) \simeq \mathfrak{bms}_3$. Also for Carrollian spacetimes describing null hypersurfaces embedded in pseudo-Riemannian spacetimes and for those obtained by the ultrarelativistic limit, the Weyl transformation has $z=1$ \cite{Ciambelli:2018wre,Ciambelli:2018xat}.

In the zweibein formulation, the Carroll-Weyl transformation defined in \cite{Ciambelli:2019lap} for general $z$, can be written as
\begin{equation}\label{Carroll-Weyl}
	e^0_{\mu} \rightarrow e^{z\Omega(\tau,\sigma)}e^0_{\mu}, \quad h_{\mu\nu} \rightarrow e^{2\Omega(\tau,\sigma)}h_{\mu\nu},
\end{equation}
where $h_{\mu\nu} = e^1_{\mu}e^1_{\nu}$. For $z=1$, this Carroll-Weyl transformation gives the BMS-Weyl transformation \eqref{BMS-Weyl}.

\newpage



\begin{thebibliography}{99}
	
	
\bibitem{Harvey:2005it}
J.~A.~Harvey,
``TASI 2003 lectures on anomalies,''
[arXiv:hep-th/0509097 [hep-th]].

\bibitem{Bilal:2008qx}
A.~Bilal,
``Lectures on Anomalies,''
[arXiv:0802.0634 [hep-th]].

\bibitem{Arav:2014goa}
I.~Arav, S.~Chapman and Y.~Oz,
``Lifshitz Scale Anomalies,''
JHEP \textbf{02}, 078 (2015)
doi:10.1007/JHEP02(2015)078
[arXiv:1410.5831 [hep-th]].


\bibitem{Jensen:2014hqa}
K.~Jensen,
``Anomalies for Galilean fields,''
SciPost Phys. \textbf{5} (2018) no.1, 005
doi:10.21468/SciPostPhys.5.1.005
[arXiv:1412.7750 [hep-th]].

\bibitem{Auzzi:2015fgg}
R.~Auzzi, S.~Baiguera and G.~Nardelli,
``On Newton-Cartan trace anomalies,''
JHEP \textbf{02}, 003 (2016)
[erratum: JHEP \textbf{02}, 177 (2016)]
doi:10.1007/JHEP02(2016)177
[arXiv:1511.08150 [hep-th]].

\bibitem{Arav:2016xjc}
I.~Arav, S.~Chapman and Y.~Oz,
``Non-Relativistic Scale Anomalies,''
JHEP \textbf{06}, 158 (2016)
doi:10.1007/JHEP06(2016)158
[arXiv:1601.06795 [hep-th]].

\bibitem{Pal:2016rpz}
S.~Pal and B.~Grinstein,
``Weyl Consistency Conditions in Non-Relativistic Quantum Field Theory,''
JHEP \textbf{12} (2016), 012
doi:10.1007/JHEP12(2016)012
[arXiv:1605.02748 [hep-th]].


\bibitem{Auzzi:2016lrq}
R.~Auzzi, S.~Baiguera, F.~Filippini and G.~Nardelli,
``On Newton-Cartan local renormalization group and anomalies,''
JHEP \textbf{11}, 163 (2016)
doi:10.1007/JHEP11(2016)163
[arXiv:1610.00123 [hep-th]].

\bibitem{Auzzi:2017jry}
R.~Auzzi, S.~Baiguera and G.~Nardelli,
``Trace anomaly for non-relativistic fermions,''
JHEP \textbf{08}, 042 (2017)
doi:10.1007/JHEP08(2017)042
[arXiv:1705.02229 [hep-th]].

\bibitem{Jensen:2017tnb}
K.~Jensen,
``Locality and anomalies in warped conformal field theory,''
JHEP \textbf{12}, 111 (2017)
doi:10.1007/JHEP12(2017)111
[arXiv:1710.11626 [hep-th]].

\bibitem{LevyLeblond}
J.~M.~L\'evy-Leblond, ``Une nouvelle limite non-relativiste du groupe de Poincar\'e,'' Annales de I'ILHP Physique th\'eorique, Volume \textbf{3}, Issue 1 (1965).


\bibitem{Bondi:1962px}
H.~Bondi, M.~G.~J.~van der Burg and A.~W.~K.~Metzner,
``Gravitational waves in general relativity. 7. Waves from axisymmetric isolated systems,''
Proc. Roy. Soc. Lond. A \textbf{269} (1962), 21-52
doi:10.1098/rspa.1962.0161

\bibitem{Sachs:1962zza}
R.~Sachs,
``Asymptotic symmetries in gravitational theory,''
Phys. Rev. \textbf{128} (1962), 2851-2864
doi:10.1103/PhysRev.128.2851





\bibitem{Barnich:2006av}
G.~Barnich and G.~Compere,
``Classical central extension for asymptotic symmetries at null infinity in three spacetime dimensions,''
Class. Quant. Grav. \textbf{24} (2007), F15-F23
doi:10.1088/0264-9381/24/5/F01
[arXiv:gr-qc/0610130 [gr-qc]].

\bibitem{Strominger:2013jfa}
A.~Strominger,
``On BMS Invariance of Gravitational Scattering,''
JHEP \textbf{07} (2014), 152
doi:10.1007/JHEP07(2014)152
[arXiv:1312.2229 [hep-th]].


\bibitem{Strominger:2017zoo}
A.~Strominger,
``Lectures on the Infrared Structure of Gravity and Gauge Theory,''
[arXiv:1703.05448 [hep-th]].

\bibitem{Duval:2014uoa}
C.~Duval, G.~W.~Gibbons, P.~A.~Horvathy and P.~M.~Zhang,
``Carroll versus Newton and Galilei: two dual non-Einsteinian concepts of time,''
Class. Quant. Grav. \textbf{31}, 085016 (2014)
doi:10.1088/0264-9381/31/8/085016
[arXiv:1402.0657 [gr-qc]].

\bibitem{Duval:2014uva}
C.~Duval, G.~W.~Gibbons and P.~A.~Horvathy,
``Conformal Carroll groups and BMS symmetry,''
Class. Quant. Grav. \textbf{31}, 092001 (2014)
doi:10.1088/0264-9381/31/9/092001
[arXiv:1402.5894 [gr-qc]].

\bibitem{Duval:2014lpa}
C.~Duval, G.~W.~Gibbons and P.~A.~Horvathy,
``Conformal Carroll groups,''
J. Phys. A \textbf{47}, no.33, 335204 (2014)
doi:10.1088/1751-8113/47/33/335204
[arXiv:1403.4213 [hep-th]].


\bibitem{Bagchi:2010zz}
A.~Bagchi,
``Correspondence between Asymptotically Flat Spacetimes and Nonrelativistic Conformal Field Theories,''
Phys. Rev. Lett. \textbf{105}, 171601 (2010)
doi:10.1103/PhysRevLett.105.171601
[arXiv:1006.3354 [hep-th]].

\bibitem{tHooft:1993dmi}
G.~'t Hooft,
``Dimensional reduction in quantum gravity,''
Conf. Proc. C \textbf{930308}, 284-296 (1993)
[arXiv:gr-qc/9310026 [gr-qc]].

\bibitem{Susskind:1994vu}
L.~Susskind,
``The World as a hologram,''
J. Math. Phys. \textbf{36} (1995), 6377-6396
doi:10.1063/1.531249
[arXiv:hep-th/9409089 [hep-th]].

\bibitem{Maldacena:1997re}
J.~M.~Maldacena,
``The Large N limit of superconformal field theories and supergravity,''
Adv. Theor. Math. Phys. \textbf{2} (1998), 231-252
doi:10.1023/A:1026654312961
[arXiv:hep-th/9711200 [hep-th]].

\bibitem{deBoer:2003vf}
J.~de Boer and S.~N.~Solodukhin,
``A Holographic reduction of Minkowski space-time,''
Nucl. Phys. B \textbf{665} (2003), 545-593
doi:10.1016/S0550-3213(03)00494-2
[arXiv:hep-th/0303006 [hep-th]].

\bibitem{Barnich:2010eb}
G.~Barnich and C.~Troessaert,
``Aspects of the BMS/CFT correspondence,''
JHEP \textbf{05} (2010), 062
doi:10.1007/JHEP05(2010)062
[arXiv:1001.1541 [hep-th]].

\bibitem{Barnich:2009se}
G.~Barnich and C.~Troessaert,
``Symmetries of asymptotically flat 4 dimensional spacetimes at null infinity revisited,''
Phys. Rev. Lett. \textbf{105} (2010), 111103
doi:10.1103/PhysRevLett.105.111103
[arXiv:0909.2617 [gr-qc]].


\bibitem{Bagchi:2012cy}
A.~Bagchi and R.~Fareghbal,
``BMS/GCA Redux: Towards Flatspace Holography from Non-Relativistic Symmetries,''
JHEP \textbf{10} (2012), 092
doi:10.1007/JHEP10(2012)092
[arXiv:1203.5795 [hep-th]].

\bibitem{Bagchi:2012xr}
A.~Bagchi, S.~Detournay, R.~Fareghbal and J.~Sim\'on,
``Holography of 3D Flat Cosmological Horizons,''
Phys. Rev. Lett. \textbf{110} (2013) no.14, 141302
doi:10.1103/PhysRevLett.110.141302
[arXiv:1208.4372 [hep-th]].

\bibitem{Bagchi:2012yk}
A.~Bagchi, S.~Detournay and D.~Grumiller,
``Flat-Space Chiral Gravity,''
Phys. Rev. Lett. \textbf{109}, 151301 (2012)
doi:10.1103/PhysRevLett.109.151301
[arXiv:1208.1658 [hep-th]].

\bibitem{Barnich:2012xq}
G.~Barnich,
``Entropy of three-dimensional asymptotically flat cosmological solutions,''
JHEP \textbf{10}, 095 (2012)
doi:10.1007/JHEP10(2012)095
[arXiv:1208.4371 [hep-th]].

\bibitem{Barnich:2012rz}
G.~Barnich, A.~Gomberoff and H.~A.~Gonz\'alez,
``Three-dimensional Bondi-Metzner-Sachs invariant two-dimensional field theories as the flat limit of Liouville theory,''
Phys. Rev. D \textbf{87}, no.12, 124032 (2013)
doi:10.1103/PhysRevD.87.124032
[arXiv:1210.0731 [hep-th]].

\bibitem{Bagchi:2013lma}
A.~Bagchi, S.~Detournay, D.~Grumiller and J.~Simon,
``Cosmic Evolution from Phase Transition of Three-Dimensional Flat Space,''
Phys. Rev. Lett. \textbf{111}, no.18, 181301 (2013)
doi:10.1103/PhysRevLett.111.181301
[arXiv:1305.2919 [hep-th]].

\bibitem{Detournay:2014fva}
S.~Detournay, D.~Grumiller, F.~Sch\"oller and J.~Sim\'on,
``Variational principle and one-point functions in three-dimensional flat space Einstein gravity,''
Phys. Rev. D \textbf{89}, no.8, 084061 (2014)
doi:10.1103/PhysRevD.89.084061
[arXiv:1402.3687 [hep-th]].

\bibitem{Afshar:2013vka}
H.~Afshar, A.~Bagchi, R.~Fareghbal, D.~Grumiller and J.~Rosseel,
``Spin-3 Gravity in Three-Dimensional Flat Space,''
Phys. Rev. Lett. \textbf{111}, no.12, 121603 (2013)
doi:10.1103/PhysRevLett.111.121603
[arXiv:1307.4768 [hep-th]].

\bibitem{Gonzalez:2013oaa}
H.~A.~Gonzalez, J.~Matulich, M.~Pino and R.~Troncoso,
``Asymptotically flat spacetimes in three-dimensional higher spin gravity,''
JHEP \textbf{09}, 016 (2013)
doi:10.1007/JHEP09(2013)016
[arXiv:1307.5651 [hep-th]].

\bibitem{Bagchi:2014iea}
A.~Bagchi, R.~Basu, D.~Grumiller and M.~Riegler,
``Entanglement entropy in Galilean conformal field theories and flat holography,''
Phys. Rev. Lett. \textbf{114}, no.11, 111602 (2015)
doi:10.1103/PhysRevLett.114.111602
[arXiv:1410.4089 [hep-th]].


\bibitem{Bagchi:2015wna}
A.~Bagchi, D.~Grumiller and W.~Merbis,
``Stress tensor correlators in three-dimensional gravity,''
Phys. Rev. D \textbf{93}, no.6, 061502 (2016)
doi:10.1103/PhysRevD.93.061502
[arXiv:1507.05620 [hep-th]].

\bibitem{Bagchi:2016geg}
A.~Bagchi, M.~Gary and Zodinmawia,
``Bondi-Metzner-Sachs bootstrap,''
Phys. Rev. D \textbf{96}, no.2, 025007 (2017)
doi:10.1103/PhysRevD.96.025007
[arXiv:1612.01730 [hep-th]].

\bibitem{Barnich:2015mui}
G.~Barnich, H.~A.~Gonzalez, A.~Maloney and B.~Oblak,
``One-loop partition function of three-dimensional flat gravity,''
JHEP \textbf{04}, 178 (2015)
doi:10.1007/JHEP04(2015)178
[arXiv:1502.06185 [hep-th]].

\bibitem{Barnich:2014cwa}
G.~Barnich, L.~Donnay, J.~Matulich and R.~Troncoso,
``Asymptotic symmetries and dynamics of three-dimensional flat supergravity,''
JHEP \textbf{08}, 071 (2014)
doi:10.1007/JHEP08(2014)071
[arXiv:1407.4275 [hep-th]].


\bibitem{Fareghbal:2014qga}
R.~Fareghbal and A.~Naseh,
``Aspects of Flat/CCFT Correspondence,''
Class. Quant. Grav. \textbf{32}, 135013 (2015)
doi:10.1088/0264-9381/32/13/135013
[arXiv:1408.6932 [hep-th]].

\bibitem{Jiang:2017ecm}
H.~Jiang, W.~Song and Q.~Wen,
``Entanglement Entropy in Flat Holography,''
JHEP \textbf{07}, 142 (2017)
doi:10.1007/JHEP07(2017)142
[arXiv:1706.07552 [hep-th]].

\bibitem{Hijano:2017eii}
E.~Hijano and C.~Rabideau,
``Holographic entanglement and Poincar\'e blocks in three-dimensional flat space,''
JHEP \textbf{05}, 068 (2018)
doi:10.1007/JHEP05(2018)068
[arXiv:1712.07131 [hep-th]].

\bibitem{Hijano:2018nhq}
E.~Hijano,
``Semi-classical BMS$_{3}$ blocks and flat holography,''
JHEP \textbf{10}, 044 (2018)
doi:10.1007/JHEP10(2018)044
[arXiv:1805.00949 [hep-th]].

\bibitem{Grumiller:2019xna}
D.~Grumiller, P.~Parekh and M.~Riegler,
``Local quantum energy conditions in non-Lorentz-invariant quantum field theories,''
Phys. Rev. Lett. \textbf{123}, no.12, 121602 (2019)
doi:10.1103/PhysRevLett.123.121602
[arXiv:1907.06650 [hep-th]].

\bibitem{Hartong:2015usd}
J.~Hartong,
``Holographic Reconstruction of 3D Flat Space-Time,''
JHEP \textbf{10}, 104 (2016)
doi:10.1007/JHEP10(2016)104
[arXiv:1511.01387 [hep-th]].


\bibitem{Hartong:2015xda}
J.~Hartong,
``Gauging the Carroll Algebra and Ultra-Relativistic Gravity,''
JHEP \textbf{08}, 069 (2015)
doi:10.1007/JHEP08(2015)069
[arXiv:1505.05011 [hep-th]].


\bibitem{Ciambelli:2018wre}
L.~Ciambelli, C.~Marteau, A.~C.~Petkou, P.~M.~Petropoulos and K.~Siampos,
``Flat holography and Carrollian fluids,''
JHEP \textbf{07}, 165 (2018)
doi:10.1007/JHEP07(2018)165
[arXiv:1802.06809 [hep-th]].

\bibitem{Donnay:2019jiz}
L.~Donnay and C.~Marteau,
``Carrollian Physics at the Black Hole Horizon,''
Class. Quant. Grav. \textbf{36}, no.16, 165002 (2019)
doi:10.1088/1361-6382/ab2fd5
[arXiv:1903.09654 [hep-th]].


\bibitem{Hawking:2016msc}
S.~W.~Hawking, M.~J.~Perry and A.~Strominger,
``Soft Hair on Black Holes,''
Phys. Rev. Lett. \textbf{116} (2016) no.23, 231301
doi:10.1103/PhysRevLett.116.231301
[arXiv:1601.00921 [hep-th]].


\bibitem{Carlip:2017xne}
S.~Carlip,
``Black Hole Entropy from Bondi-Metzner-Sachs Symmetry at the Horizon,''
Phys. Rev. Lett. \textbf{120} (2018) no.10, 101301
doi:10.1103/PhysRevLett.120.101301
[arXiv:1702.04439 [gr-qc]].

\bibitem{Carlip:2019dbu}
S.~Carlip,
``Near-horizon Bondi-Metzner-Sachs symmetry, dimensional reduction, and black hole entropy,''
Phys. Rev. D \textbf{101} (2020) no.4, 046002
doi:10.1103/PhysRevD.101.046002
[arXiv:1910.01762 [hep-th]].

\bibitem{Gross:1987kza}
D.~J.~Gross and P.~F.~Mende,
``The High-Energy Behavior of String Scattering Amplitudes,''
Phys. Lett. B \textbf{197} (1987), 129-134
doi:10.1016/0370-2693(87)90355-8

\bibitem{Gross:1987ar}
D.~J.~Gross and P.~F.~Mende,
``String Theory Beyond the Planck Scale,''
Nucl. Phys. B \textbf{303} (1988), 407-454
doi:10.1016/0550-3213(88)90390-2

\bibitem{Isberg:1993av}
J.~Isberg, U.~Lindstrom, B.~Sundborg and G.~Theodoridis,
``Classical and quantized tensionless strings,''
Nucl. Phys. B \textbf{411} (1994), 122-156
doi:10.1016/0550-3213(94)90056-6
[arXiv:hep-th/9307108 [hep-th]].

\bibitem{Bagchi:2013bga}
A.~Bagchi,
``Tensionless Strings and Galilean Conformal Algebra,''
JHEP \textbf{05} (2013), 141
doi:10.1007/JHEP05(2013)141
[arXiv:1303.0291 [hep-th]].

\bibitem{Bagchi:2015nca}
A.~Bagchi, S.~Chakrabortty and P.~Parekh,
``Tensionless Strings from Worldsheet Symmetries,''
JHEP \textbf{01} (2016), 158
doi:10.1007/JHEP01(2016)158
[arXiv:1507.04361 [hep-th]].

\bibitem{Bagchi:2020ats}
A.~Bagchi, A.~Banerjee and S.~Chakrabortty,
``Rindler Physics on the String Worldsheet,''
Phys. Rev. Lett. \textbf{126}, no.3, 031601 (2021)
doi:10.1103/PhysRevLett.126.031601
[arXiv:2009.01408 [hep-th]].


\bibitem{Bagchi:2020fpr}
A.~Bagchi, A.~Banerjee, S.~Chakrabortty, S.~Dutta and P.~Parekh,
``A tale of three \textemdash{} tensionless strings and vacuum structure,''
JHEP \textbf{04}, 061 (2020)
doi:10.1007/JHEP04(2020)061
[arXiv:2001.00354 [hep-th]].



\bibitem{Bagchi:2019cay}
A.~Bagchi, A.~Banerjee and P.~Parekh,
``Tensionless Path from Closed to Open Strings,''
Phys. Rev. Lett. \textbf{123}, no.11, 111601 (2019)
doi:10.1103/PhysRevLett.123.111601
[arXiv:1905.11732 [hep-th]].



\bibitem{Bagchi:2013qva}
A.~Bagchi and R.~Basu,
``3D Flat Holography: Entropy and Logarithmic Corrections,''
JHEP \textbf{03} (2014), 020
doi:10.1007/JHEP03(2014)020
[arXiv:1312.5748 [hep-th]].

\bibitem{Henningson:1998gx}
M.~Henningson and K.~Skenderis,
``The Holographic Weyl anomaly,''
JHEP \textbf{07} (1998), 023
doi:10.1088/1126-6708/1998/07/023
[arXiv:hep-th/9806087 [hep-th]].

\bibitem{Duff:1993wm}
M.~J.~Duff,
``Twenty years of the Weyl anomaly,''
Class. Quant. Grav. \textbf{11} (1994), 1387-1404
doi:10.1088/0264-9381/11/6/004
[arXiv:hep-th/9308075 [hep-th]].

\bibitem{Tong:2009np}
D.~Tong,
``String Theory,''
[arXiv:0908.0333 [hep-th]].

\bibitem{Bagchi:2010vw}
A.~Bagchi,
``Topologically Massive Gravity and Galilean Conformal Algebra: A Study of Correlation Functions,''
JHEP \textbf{02}, 091 (2011)
doi:10.1007/JHEP02(2011)091
[arXiv:1012.3316 [hep-th]].


\bibitem{Basu:2015evh}
R.~Basu and M.~Riegler,
``Wilson Lines and Holographic Entanglement Entropy in Galilean Conformal Field Theories,''
Phys. Rev. D \textbf{93}, no.4, 045003 (2016)
doi:10.1103/PhysRevD.93.045003
[arXiv:1511.08662 [hep-th]].

\bibitem{Bagchi:2009my}
A.~Bagchi and R.~Gopakumar,
``Galilean Conformal Algebras and AdS/CFT,''
JHEP \textbf{07}, 037 (2009)
doi:10.1088/1126-6708/2009/07/037
[arXiv:0902.1385 [hep-th]].

\bibitem{Duval:2009vt}
C.~Duval and P.~A.~Horvathy,
``Non-relativistic conformal symmetries and Newton-Cartan structures,''
J. Phys. A \textbf{42}, 465206 (2009)
doi:10.1088/1751-8113/42/46/465206
[arXiv:0904.0531 [math-ph]].

\bibitem{Kraus:2005zm}
P.~Kraus and F.~Larsen,
``Holographic gravitational anomalies,''
JHEP \textbf{01}, 022 (2006)
doi:10.1088/1126-6708/2006/01/022
[arXiv:hep-th/0508218 [hep-th]].

\bibitem{Skenderis:2009nt}
K.~Skenderis, M.~Taylor and B.~C.~van Rees,
``Topologically Massive Gravity and the AdS/CFT Correspondence,''
JHEP \textbf{09}, 045 (2009)
doi:10.1088/1126-6708/2009/09/045
[arXiv:0906.4926 [hep-th]].





\bibitem{Bagchi:2009pe}
A.~Bagchi, R.~Gopakumar, I.~Mandal and A.~Miwa,
``GCA in 2d,''
JHEP \textbf{08}, 004 (2010)
doi:10.1007/JHEP08(2010)004
[arXiv:0912.1090 [hep-th]].

\bibitem{Deser:1982vy}
S.~Deser, R.~Jackiw and S.~Templeton,
``Three-Dimensional Massive Gauge Theories,''
Phys. Rev. Lett. \textbf{48}, 975-978 (1982)
doi:10.1103/PhysRevLett.48.975

\bibitem{Bagchi:2013hja}
A.~Bagchi and D.~Grumiller,
``Holograms of flat space,''
Int. J. Mod. Phys. D \textbf{22}, 1342003 (2013)
doi:10.1142/S0218271813420030

\bibitem{Grumiller:2015xaa}
D.~Grumiller and W.~Merbis,
``Free energy of topologically massive gravity and flat space holography,''
Springer Proc. Phys. \textbf{208}, 95-103 (2018)
doi:10.1007/978-3-319-94256-8\_10
[arXiv:1509.08505 [hep-th]].

\bibitem{Bagchi:2018ryy}
A.~Bagchi, R.~Basu, S.~Detournay and P.~Parekh,
``Flatspace Chiral Supergravity,''
Phys. Rev. D \textbf{97}, no.10, 106020 (2018)
doi:10.1103/PhysRevD.97.106020
[arXiv:1801.03245 [hep-th]].


\bibitem{Barnich:2013yka}
G.~Barnich and H.~A.~Gonzalez,
``Dual dynamics of three dimensional asymptotically flat Einstein gravity at null infinity,''
JHEP \textbf{05}, 016 (2013)
doi:10.1007/JHEP05(2013)016
[arXiv:1303.1075 [hep-th]].



\bibitem{Bergshoeff:2017btm}
E.~Bergshoeff, J.~Gomis, B.~Rollier, J.~Rosseel and T.~ter Veldhuis,
``Carroll versus Galilei Gravity,''
JHEP \textbf{03}, 165 (2017)
doi:10.1007/JHEP03(2017)165
[arXiv:1701.06156 [hep-th]].

\bibitem{inprogress}
A.~Bagchi, A.~Banerjee, S.~Dutta, K.~Kolekar, P.~Sharma,
Work in progress. 

\bibitem{Ciambelli:2019lap}
L.~Ciambelli, R.~G.~Leigh, C.~Marteau and P.~M.~Petropoulos,
``Carroll Structures, Null Geometry and Conformal Isometries,''
Phys. Rev. D \textbf{100}, no.4, 046010 (2019)
doi:10.1103/PhysRevD.100.046010
[arXiv:1905.02221 [hep-th]].

\bibitem{Ciambelli:2018xat}
L.~Ciambelli, C.~Marteau, A.~C.~Petkou, P.~M.~Petropoulos and K.~Siampos,
``Covariant Galilean versus Carrollian hydrodynamics from relativistic fluids,''
Class. Quant. Grav. \textbf{35}, no.16, 165001 (2018)
doi:10.1088/1361-6382/aacf1a
[arXiv:1802.05286 [hep-th]].

\bibitem{Gupta:2020dtl}
N.~Gupta and N.~V.~Suryanarayana,
``Constructing Carrollian CFTs,''
JHEP \textbf{03}, 194 (2021)
doi:10.1007/JHEP03(2021)194
[arXiv:2001.03056 [hep-th]].



%
%



\end{thebibliography}
\end{document}